\documentclass[11pt]{article}
\usepackage[T1]{fontenc}
\usepackage[utf8]{inputenc}
\usepackage{lmodern}
\usepackage{upgreek}
\usepackage{jheppub}
\usepackage{subcaption}
\usepackage{ragged2e}
\usepackage[normalem]{ulem}
 
\usepackage[T1]{fontenc}   
\usepackage[utf8]{inputenc}
 
\usepackage{amsmath,amssymb,amsfonts,dsfont,mathrsfs,amsthm,mathtools,bbm}
\usepackage{braket}
\usepackage{xcolor} 
 
\definecolor{rossocorsa}{rgb}{0.83, 0.0, 0.0}
\definecolor{bleudefrance}{rgb}{0.19, 0.55, 0.91}

\usepackage{hyperref}
\hypersetup{
    colorlinks,
    citecolor=bleudefrance,
    filecolor=bleudefrance,
    linkcolor=rossocorsa,
    urlcolor=bleudefrance
}

\newcommand{\stkout}[1]{\ifmmode\text{\sout{\ensuremath{#1}}}\else\sout{#1}\fi}

\newcommand{\be}{\begin{equation}}
\newcommand{\ee}{\end{equation}}
\newcommand{\bea}{\begin{eqnarray}}
\newcommand{\eea}{\end{eqnarray}}
\newcommand{\beas}{\begin{eqnarray*}}
\newcommand{\eeas}{\end{eqnarray*}}

\newcommand{\Area}{\operatorname{Area}_\Gamma(\Sigma_A)}
\newcommand{\Ls}{L}
\newcommand{\dS}{\text{dS}}
\newcommand{\GN}{G_{\text N}}
\newcommand{\diff}{\text{d}}
\newcommand{\e}[1]{\text{e}^{#1}}

\newcommand{\tE}{t_\text{E}}
\newcommand{\iu}{\text i}

\newcommand{\tauUV}{\uptau_{\mathrm{UV}}}

\begin{document}
\noindent\flushright
{YITP-26-138} 
\noindent\flushleft\justifying
\vspace{-2cm}
\title{\huge Holomorphic extremal surfaces for pseudoentropy in dS/CFT}
 
\author[a]{Giorgos Anastasiou,}
\author[b]{Ignacio J. Araya,}
\author[c, d, e]{Avijit Das,}
\author[e]{Javier Moreno}

\affiliation[a]{Departamento de Ciencias, Facultad de Artes Liberales, Universidad Adolfo Ibañez,\\ Av. Diagonal Las Torres, 2640 Peñalolen, Chile \vspace{0.1cm}}
\affiliation[b]{Departamento de Física y Astronomía, Facultad de Ciencias Exactas, Universidad Andres Bello,\\ Sazié 2212, Piso 7, Santiago, Chile \vspace{0.1cm}}
\affiliation[c]{Instituto de Ciencias Exactas y Naturales, Universidad Arturo Prat,\\
Playa Brava 3256, 1111346, Iquique, Chile \vspace{0.1cm}}
\affiliation[d]{Facultad de Ciencias, Universidad Arturo Prat,\\
Avenida Arturo Prat Chacón 2120, 1110939, Iquique, Chile \vspace{0.1cm}}
\affiliation[e]{Center for Gravitational Physics and Quantum Information,\\
Yukawa Institute for Theoretical Physics, Kyoto University,\\
Kitashirakawa Oiwakecho, Sakyo-ku, Kyoto 606-8502, Japan
\vspace{0.1cm}
}
 
\vspace{0.3cm}

\abstract{
We formulate the extremization problem for codimension-two hypersurfaces in de Sitter space in holomorphic coordinates, with real anchoring conditions at future infinity. The resulting equation ($K^I=0$) describes complex bulk realizations extending the usual piecewise Hartle--Hawking construction. We compute holographic pseudoentropy for harmonic deformations of spherical regions of arbitrary size and dimension, confirming its universal expansion through second order in the deformation parameter. The result is independent of the integration contour within each homology class on the complexified extremal surface. Inequivalent classes, labeled by an integer associated with a discrete global-coordinate isometry, differ by a sign of the universal term in odd boundary dimensions and by real finite shifts in even dimensions. For conical defects in dS$_3$ parametrized by a Schwarzschild-dS mass, we also verify the first law of pseudoentropy and the persistence of contour
degeneracy within each homology class.}
 
\maketitle

\section{Introduction}

In the holographic framework, pseudoentropy emerges as a prominent quantity within the de Sitter/conformal field theory (dS/CFT) correspondence~\cite{Strominger:2001pn,Witten:2001kn,Maldacena:2002vr}, where gravity in a dS spacetime is conjectured to be dual to a CFT defined on the boundary at future time infinity. Unlike its anti-de Sitter counterpart (AdS/CFT)~\cite{Maldacena:1997re,Gubser:1998bc,Witten:1998qj}, the CFT dual to dS space is nonunitary~\cite{Narayan:2015vda,Narayan:2015oka,Sato:2015tta,Narayan:2016xwq}, and only a limited number of explicit realizations of the correspondence have been constructed so far~\cite{Anninos:2011ui,Ng:2012xp,Das:2012dt,Chang:2013afa,Mollabashi:2021xsd,Hikida:2022ltr}. In this setting, the holographic entanglement entropy, obtained through the Ryu--Takayanagi (RT) formula~\cite{Ryu:2006bv,Ryu:2006ef,Hubeny:2007xt}, generally acquires complex values, so that the natural quantity computed by the extremal-surface prescription is not the entanglement entropy itself but its complex generalization, the pseudoentropy~\cite{Nakata:2020luh}, originally introduced more broadly to describe non-Hermitian quantum systems~\cite{Hatano_1996,Bender:2007nj,Castro-Alvaredo:2017udm,Couvreur:2016mbr,Herviou:2019yfb,Chang:2019jcj,Bergholtz:2019deh,Jian:2020byi,Ashida:2020dkc,Bender:2023cem,Caputa:2024gve,Shimizu:2025kse,Harper:2025lav,Kawamoto:2025oko,Maeda:2026awj,Fujiki:2026kca}. In the original holographic formulation of~\cite{Doi:2022iyj,Doi:2023zaf}---see also~\cite{Narayan:2022afv,Narayan:2023zen}, the geometric prescription that reproduces this structure in a suitable extremal surface corresponds to the Hartle--Hawking (HH) state~\cite{Hartle:1983ai,Gibbons:1977mu}, for which the extremal surface traverses both the Euclidean and Lorentzian regions of the spacetime. As a result, its area acquires a real contribution from the spacelike (Euclidean) segment and an imaginary contribution from the timelike (Lorentzian) segment, yielding the complex pseudoentropy expected of the dual nonunitary CFT.
 
A closely related quantity, timelike entanglement entropy, has recently attracted considerable attention as a measure of quantum correlations across timelike-separated regions~\cite{Doi:2022iyj,Doi:2023zaf}---see~\cite{Jiang:2023ffu,Chu:2023zah,Foligno:2023dih,Das:2023yyl,Jiang:2023loq,Grieninger:2023knz,Guo:2024lrr,Xu:2024yvf,Afrasiar:2024lsi,Nunez:2025ppd,Nunez:2025gxq,Zhao:2025zgm,Alishahiha:2025xml}, among others. In its original holographic description through AdS/CFT, the extremal surface associated to timelike entanglement entropy is likewise composed of timelike and spacelike segments; under Wick rotation, these are mapped to their pseudoentropy counterparts with the two segments exchanging character---the timelike segment in AdS maps to the spacelike segment in dS, and vice versa.
 
While in their original formulations pseudoentropy and timelike entanglement entropy were computed via piecewise prescriptions for spherical subregions, yielding sensible results, it was noted that this construction carried severe limitations. In particular: i) certain subregions in both setups did not allow for a clear interpretation, such as the strip or small spherical subregions\footnote{As we will mention later, the case of holographic pseudoentropy for small perturbations around the spherical subregion~\cite{Anastasiou:2025rvz,Anastasiou:2026bbf} (of ``unit'' radius) also bypasses this problem.}~\cite{Doi:2023zaf,Narayan:2026wzp}; and ii) it proved impossible to formulate a holographic first law for either quantity~\cite{Fujiki:2025rtx,Li:2025tud,Xiao:2026nvu}. In the case of timelike entanglement entropy in AdS/CFT, these limitations motivate replacing the piecewise construction by genuinely complex, boundary-anchored codimension-two extremal surfaces in the complexified bulk~\cite{Heller:2024whi}. This proposal has subsequently been formulated more systematically as an analytic continuation of the full set of codimension-two extremal surfaces satisfying the homology constraint, together with a prescription for selecting the dominant complex     saddle~\cite{Heller:2025kvp,Bernamonti:2026pxo}. A closely related problem of selecting admissible complex extremal surfaces arises in both timelike AdS/CFT and dS/CFT, where additional conditions on the associated complex geometry may constrain the allowed integration contours~\cite{Guo:2026vkl}.\footnote{See also~\cite{Kontsevich:2021dmb,Witten:2021nzp,Guo:2026vkl} for discussions on the complex saddles in the context of the Euclidean path integral.} These developments provide a direct motivation for treating holographic pseudoentropy in the same language: rather than constructing the surface by joining distinct real Lorentzian and Euclidean segments, the extremization problem can be formulated intrinsically in the complexified dS bulk.

More generally, complex codimension-two extremal surfaces have long appeared as extensions of the RT prescription once the relevant holographic observable cannot be represented by a real extremal surface. In asymptotically AdS black-hole geometries, complex codimension-two extremal surfaces and their different analytic branches were already considered as possible contributions to holographic entropy~\cite{Fischetti:2014zja}. On the dS side, complex codimension-two extremal surfaces anchored at future infinity arise naturally in dS/CFT, both for pure dS and for deformations of the background~\cite{Narayan:2015vda,Narayan:2015oka}. More recently, the no-boundary prescription has made explicit that the corresponding areas may be formulated through integration contours in complexified time, with different contour representatives becoming equivalent whenever they can be continuously deformed into one another without crossing an obstruction~\cite{Narayan:2022afv,Narayan:2023zen,Narayan:2026wzp}. These developments suggest that complex bulk geometry is not merely a technical continuation of a real prescription, but can constitute the natural domain in which the codimension-two extremization problem is defined.

Based on these considerations, in Sec.~\ref{sec:hes} we develop a framework for formulating the extremization problem for holographic pseudoentropy directly in complexified dS coordinates, treating the metric as the holomorphic extension of the global Lorentzian dS line element. In this formulation, the metric reduces to that of Lorentzian dS when the coordinates are real, and to that of Euclidean dS when the global time is purely imaginary, while also allowing for generic complex values of all coordinates. The embedding of the extremal surface, subject to real boundary conditions at future infinity, can then be specified directly in terms of these holomorphic coordinates. We then illustrate our holomorphic extremization prescription by computing the holographic pseudoentropy associated with a spherical subregion located at asymptotic future infinity, and show agreement with the expected results for arbitrary subregion size in dS$_4$/CFT$_3$.  Next, we turn to harmonic deformations of a spherical entangling surface in the same dimension, recovering the results of~\cite{Anastasiou:2025rvz,Anastasiou:2026bbf} for perturbations around the ``unit'' radius corresponding to the maximal entangling region and extending it to arbitrary size. In Sec.~\ref{sec:hdim}, we extend our results to arbitrary higher dimensions, and explain how the integer labeling the turning points of the complex surface enters the result. In Sec.~\ref{sec:ds3-conical-defect} we consider conical defects in dS$_3$ as an excited-state example, before concluding in Sec.~\ref{sec:conclusions} with some final remarks.

\section{Holomorphic extremal surfaces}\label{sec:hes}

In this section we formulate the holographic computation of pseudoentropy as
an extremization problem in complexified dS space. We first complexify
the global dS$_{d+1}$ geometry and recover the Lorentzian dS spacetime as a
real slice of it (Sec.~\ref{subsec:complexbulk}). We then construct the complex conformal
boundary and select its Euclidean real section, on which the entangling
region and the boundary data of the dual CFT are specified
(Sec.~\ref{subsec:realbdy}). On this geometry we
define the holomorphic RT prescription, organize its integration contours into
classes labeled by an integer $n$, and derive the resulting UV expansion of the
pseudoentropy (Sec.~\ref{subsec:holoRT}). Finally, we apply the prescription to
spherical entangling regions and their harmonic deformations in
dS$_4$/CFT$_3$ (Sec.~\ref{subsec:dS4example}).
 
\subsection{Complexified dS bulk spacetime}\label{subsec:complexbulk}
Let us start with the global Lorentzian dS$_{d+1}$ spacetime metric,
$\diff s^2=g_{\mu\nu}\diff x^\mu\diff x^\nu$, which reads
\begin{equation}\label{eq:GdS1} 
    \diff s^2=\Ls^2\left[-\diff\tau^2+\cosh^2\tau\left(\diff\tE^2+\sin^2\tE\diff\Omega_{d-1}^2\right)\right]\,,
\end{equation}
where $\Ls$ is the dS radius, $\tau$ is the bulk time coordinate, $\tE$ is the `Euclidean time' of the boundary\footnote{The compact Euclidean direction $\tE$ serves as one of the angular coordinates on the global boundary 
$\mathbb S^{d}$.} and $\diff\Omega_{d-1}^2=\diff\theta_1^2+\cos^2\theta_1\,\diff\Omega_{d-2}^2$ is the metric of the unit $\mathbb S^{d-1}$. The coordinates define a real chart on the Lorentzian dS spacetime,
\begin{equation}
x^\mu=(\tau,\tE,\theta_i,\phi),
\qquad
x^\mu:\text{dS}_{d+1}\rightarrow\mathbb{R}^{d+1}\,,
\end{equation}
with $\tau\in\mathbb{R},\quad
\tE\in[0,\pi],\quad \theta_i\in[-\pi/2,\pi/2],\quad
\phi\in[0,2\pi),
\quad i=1,\ldots,d-2 .$
 
We now holomorphically complexify the bulk spacetime by promoting each real coordinate to an independent complex coordinate,
\begin{equation}
\tau,\tE,\theta_i,\phi
\longrightarrow
\uptau,\mathsf{t}_{\mathrm E},\uptheta_i,\upphi\in\mathbb{C}\,.
\end{equation}

The resulting space is a complex manifold, \diff S$_{d+1}^{\mathbb C}=\mathcal{M}_{\mathbb{C}}$, 
of complex dimension $\dim_{\mathbb C}\left(\mathcal{M}_{\mathbb{C}}\right)=d+1,$ or equivalently of real dimension $\dim_{\mathbb R}\left(\mathcal{M}_{\mathbb{C}}\right)=2(d+1)$, with
\begin{equation}
\mathsf{x}^\mu =(\uptau, \mathsf{t}_{\mathrm E}, \uptheta_i,\upphi),
\qquad
\mathsf{x}^\mu :\mathrm{dS}_{d+1}^{\mathbb C}\rightarrow\mathbb{C}^{d+1}\,.
\end{equation}
 Consequently, $\diff \mathsf s^2= \mathsf{g}_{\mu \nu} \diff\mathsf{x}^\mu \diff \mathsf{x}^\nu$ defines a holomorphic symmetric $(0,2)$ tensor field $\mathsf{g}_{\mu \nu} (\mathsf{x})$ on $\mathrm{dS}_{d+1}^{\mathbb C}$. The holomorphic metric on $\mathcal{M}_{\mathbb C}$ is obtained by analytic
continuation of Eq.~\eqref{eq:GdS1},
\begin{equation}\label{eq:dScom}
\diff \mathsf s^2
=
\Ls^2
\left[
-\diff\uptau^2
+\cosh^2\uptau
\left(
\diff\mathsf{t}_{\mathrm E}^2
+\sin^2\mathsf{t}_{\mathrm E}\,
\diff\Upomega_{d-1}^2
\right)
\right]\,,
\end{equation}
where $\diff\Upomega_{d-1}^{2}$ denotes the holomorphic continuation
of the metric on the unit sphere $\mathbb S^{d-1}$. All functions appearing in the metric are understood as their holomorphic extensions to complex arguments.

The above construction is mathematically well-defined because $\dS_{d+1}$ is a
real analytic Lorentzian manifold. A classical theorem in complex geometry
states that every real analytic manifold admits a complexification, which is
locally unique up to biholomorphism. In the particular case of dS space,
this complexification admits a natural global realization since dS$_{d+1}$ can be represented as the real analytic quadric 
\begin{equation}\label{eq:quadric}
-X_0^2+X_1^2+\cdots+X_{d+1}^2
=\Ls^2 \subset\mathbb{R}^{1,d+1}\,.
\end{equation}
Its complexification is obtained by replacing the ambient space
$\mathbb{R}^{1,d+1}$ with $\mathbb{C}^{d+2}$ (see Refs.~\cite{Grauert1958OnLP,Whitney1959QuelquesPF,wells2008differential}). 
The holomorphic extension of the global coordinate chart adopted in the present work provides an explicit 
coordinate description of this global complexification.
 
The original real Lorentzian dS spacetime is recovered as the real slice of this complex manifold, $\mathrm{dS}_{d+1}\subset \mathcal{M}_{\mathbb C},$
corresponding to the locus where all complex coordinates are restricted to their real values, i.e.,
\begin{equation}
\left.
\mathcal{M}_{\mathbb C}
\right|_{\operatorname{Im}(\mathsf{x}^{\mu})=0}
=\left\{
\mathsf{x}\in\mathcal{M}_{\mathbb C}
\;\middle|\;
\operatorname{Im}(\mathsf{x}^{\mu})=0
\right\}
\simeq \dS_{d+1}\,.
\end{equation}
 
\subsection{Conformal boundary of the complexified dS spacetime}\label{subsec:realbdy}

Having defined the complexified bulk spacetime, we now construct its complex conformal boundary in a way that makes explicit the new freedom that holomorphicity gives for choosing the location of the real section of the boundary. In order to proceed, we consider the complex embedding coordinates where a global complex conformal completion can be constructed from the ambient
quadric of Eq.~\eqref{eq:quadric} by homogenization. Defining a new complex coordinate $Z^A$ by
$X^A/\Ls=Z^A/W$, one obtains the conformal completion of ${\mathcal M}_{\mathbb C}$, denoted by $\overline{\mathcal M}_{\mathbb C}$ as
\begin{equation}
    \begin{aligned}
\overline{\mathcal M}_{\mathbb C}
&=\left\{\left.Z^A \, \right| \,\eta_{AB}Z^AZ^B=W^2 \right\}\subset\mathbb{C}^{d+2}\,,\\
\mathcal D_{\mathbb C}
&=\left\{\left.Z^A \, \right|\eta_{AB}Z^AZ^B=0\right\}\,,
\end{aligned}
\end{equation}
with $A,B=0,\ldots,d+1$,
$\eta_{AB}=\operatorname{diag}(-1,1,\ldots,1)$ and $W$ a real parameter which can be rescaled.
The affine chart $W\neq0$ recovers the complexified bulk $\mathcal M_{\mathbb C}$, while
$W=0$ defines its conformal divisor $\mathcal D_{\mathbb C}$---see, for example, Ref.~\cite{Robinson:1993no}. One can then define the conformal boundary $\partial \overline{\mathcal{M}}_{\mathbb C}$ as the geometric locus of $\overline{\mathcal{M}}_{\mathbb C}$ where the condition of the conformal divisor is satisfied.\footnote{This formulation generalizes the standard construction of the conformal boundary, where conformal completion is achieved by the freedom to rescale the radius $\Ls$ asymptotically away from the origin where the dS hyperboloid tends to the mantle of the lightcone.}

The construction here discussed does not depend on the location or signature of the conformal boundary, as one could consider different real sections of $\overline{\mathcal M}_{\mathbb C}$. As it will be shown in what follows, after deciding on the signature and location of the conformal boundary, one can define a coordinate chart that is adapted to this choice.\footnote{A motivating example of the different available choices of conformal boundary reality conditions and signatures for the real sections is provided by the
Wick-rotated bulk geometry with two timelike directions and a Lorentzian
torus boundary discussed in~\cite{Fujiki:2026kca}.}

For the particular case under study, it is convenient to construct the complex
conformal boundary with its corresponding reality condition such that one recovers the
Euclidean future boundary of dS/CFT. In global coordinates, conformal
infinity occurs at an infinite value of the time coordinate. To make the
conformal compactification explicit, we introduce a holomorphic radial
chart near the asymptotic end $\operatorname{Re}\uptau\to+\infty$,
\begin{equation}\label{eq:boundary-rho}
\rho=2\e{-\uptau},
\qquad
\diff{\mathsf q}^2
=\diff\mathsf t_{\mathrm E}^{2}
+\sin^2\mathsf t_{\mathrm E}\,\diff\Upomega_{d-1}^{2}\,,
\end{equation}
where $\diff\Upomega_{d-1}^{2}$ is the holomorphic continuation of the
unit-sphere metric introduced in Sec.~\ref{subsec:complexbulk}.
Equation~\eqref{eq:dScom} then becomes
\begin{equation}\label{eq:global-conformal-pole}
\diff\mathsf s^2
=\frac{\Ls^2}{\rho^2}
\left[
-\diff\rho^2
+\left(1+\frac{\rho^2}{4}\right)^2\diff \mathsf q^2
\right]\,.
\end{equation}
Thus, the second-order pole of the complexified metric is manifest in this
chart. The transformation is defined on a punctured neighborhood of
conformal infinity; the locus $\rho=0$ is subsequently adjoined in the
conformal completion. For this coordinate choice, the complex conformal boundary near the chosen asymptotic end is the holomorphic hypersurface 
\begin{equation}
\partial \overline{\mathcal M}_{\mathbb C}
=\left.\overline{\mathcal M}_{\mathbb C}\right|_{\rho=0}\,\,\,;\,\,\, \dim_{\mathbb C}\partial \overline{\mathcal M}_{\mathbb C}=d\,.
\end{equation}
A real manifold with boundary is obtained only after a real section has been selected. The conformally rescaled metric
\begin{equation}\label{eq:complex-compactified-metric}
\diff\overline{\mathsf s}^{\,2}
=\rho^2\diff\mathsf s^2
=\Ls^2
\left[
-\diff\rho^2
+\left(1+\frac{\rho^2}{4}\right)^2\diff \mathsf q^2
\right]
\end{equation}
gives a representative holomorphic metric for
$\partial \overline{\mathcal M}_{\mathbb C}$. This is the holomorphic version
of the Penrose conformal compactification, whose defining-function
construction is reviewed in Ref.~\cite{Witten:1998qj}. Different reality conditions yield real boundary
metrics of different signatures. Their identification with the boundary
of a particular real bulk geometry additionally requires a compatible bulk
reality condition and the appropriate metric or radius phase.

For the dS/CFT setup considered here, the real bulk approaches the boundary
along $\rho>0$, corresponding to real global time tending to $+\infty$.
Accordingly, we define the future conformal boundary of
$\mathcal{M}_{\mathbb C}$ in terms of the original angular coordinates as the real section
\begin{equation}
\mathcal{I}_{\mathbb R}^{+}
=
\left\{
\mathsf{x}\in\mathcal{M}_{\mathbb C}
\;\middle|\;
\uptau\rightarrow+\infty,\;
\operatorname{Im}(\uptau)
=
\operatorname{Im}(\mathsf{t}_{\mathrm E})
=
\operatorname{Im}(\uptheta_i)
=
\operatorname{Im}(\upphi)
=0
\right\}\,.
\end{equation}
 The induced metric at the conformal boundary is therefore the same as in ordinary dS
space, such that
\begin{equation}\label{eq:boundary-real-section}
\mathcal I_{\mathbb R}^{+}
\simeq\mathbb S^d,
\qquad
\diff \mathsf{h}_{\mathbb R}^2
=\Ls^2\left(
\diff t_{\mathrm E}^{2}
+\sin^2t_{\mathrm E}\,\diff\Omega_{d-1}^{2}
\right)\,.
\end{equation}

Throughout this work, the entangling region, the CFT state and all
boundary observables are specified on this Euclidean real section
$\mathcal I_{\mathbb R}^{+}$ using the original boundary coordinates
$(\tE,\theta_i,\phi)$, whereas the corresponding codimension--two extremal
surfaces extend into $\mathcal M_{\mathbb C}$.

\begin{figure}
    \centering    \includegraphics[width=0.75\linewidth]{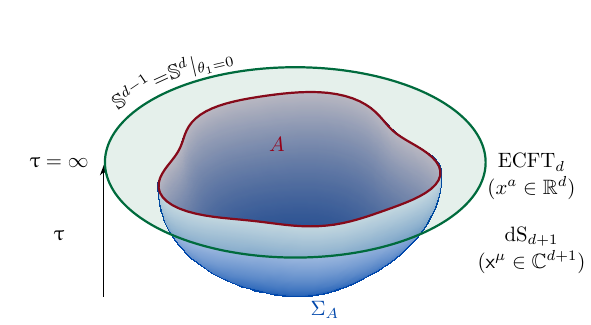}
    \caption{Representation of a complex RT surface corresponding to an arbitrary real entangling region on $\mathcal I_{\mathbb R}^{+}$. This boundary is the chosen Euclidean real section of the complex conformal boundary of $\mathcal M_{\mathbb C}$. The data of the dual nonunitary CFT are specified on this real section, while the corresponding extremal surface extends into the complexified bulk.}
    \label{fig:ComplexRT}
\end{figure}

\subsection{Holomorphic RT prescription for pseudoentropy}
\label{subsec:holoRT}

The complexified bulk of Sec.~\ref{subsec:complexbulk} and the chosen
Euclidean real section of its complex conformal boundary, defined in
Sec.~\ref{subsec:realbdy}, are the two ingredients of our proposal.
The boundary fixes the entangling region and the CFT data, while the extremal
surface that computes the pseudoentropy is allowed to extend into
$\mathcal M_{\mathbb C}$. The quantity computed holographically is the
pseudoentropy~\cite{Nakata:2020luh}. For two pure states $|\psi\rangle$ and
$|\varphi\rangle$ in $\mathcal H_{\rm tot}=\mathcal H_A\otimes\mathcal H_{\bar A}$,
it is defined in terms of the reduced transition matrix as
\begin{equation}
S(A)=-\operatorname{tr}\!\left(\tau_A\log\tau_A\right),
\qquad
\tau_A=\operatorname{tr}_{\bar A}
\left(\frac{|\psi\rangle\langle\varphi|}{\langle\varphi|\psi\rangle}\right)\,,
\end{equation}
where the two states are prepared by Euclidean path integrals of the dual CFT
on the two hemispheres of $\mathbb S^{d}$, glued along
$\bar A$~\cite{Doi:2022iyj,Doi:2023zaf}. Since the theory is nonunitary, these
preparations are not related by Hermitian conjugation, and
$\tau_A\neq\tau_A^\dagger$. Holographically, the replica trick applied to the
HH wavefunction computes $S(A)$ from the area of the
codimension-two surface $\Sigma_A$ fixed by the replica
symmetry~\cite{Lewkowycz:2013nqa}, which becomes extremal and anchored on
$\partial A$ as the number of replicas tends to one. In the original formulation~\cite{Doi:2022iyj,Doi:2023zaf}, this surface is constructed
piecewise in the HH geometry as
$\Sigma_A=\Sigma_A^{(\mathrm t)}\cup\Sigma_A^{(\mathrm s)}$.
The timelike segment $\Sigma_A^{(\mathrm t)}$ lies on the Lorentzian
section of real de Sitter space, while the spacelike segment
$\Sigma_A^{(\mathrm s)}$ lies on the Euclidean hemisphere used in the
HH preparation. Each segment is extremal in its respective
geometry, and the two are matched at the Lorentzian--Euclidean junction,
$\tau=\tau_{\mathrm E}=0$.\footnote{The Wick rotation
$\tau=-\iu\tau_{\mathrm E}$ is implied by the junction condition
$\partial_\tau t_{\mathrm E}
=\iu\,\partial_{\tau_{\mathrm E}}t_{\mathrm E}$
imposed in Refs.~\cite{Anastasiou:2025rvz,Anastasiou:2026bbf}.
With this continuation, the Lorentzian and Euclidean profiles of these
references are recovered as restrictions of the holomorphic solutions
found below on the contour class $\Gamma^{(0)}$.}

But, as discussed in the introduction, this piecewise construction does not satisfy the first law of pseudoentropy and is restricted to a very special class of subregions. To overcome these limitations, we instead take $\Sigma_A$ to be a single complex surface in
$\mathcal M_{\mathbb C}$, anchored on
$\partial A\subset\mathcal I^+_{\mathbb R}$, of which
$\Sigma_A^{(\mathrm t)}$ and $\Sigma_A^{(\mathrm s)}$ are the
restrictions to the real and imaginary $\uptau$ axes, respectively. A schematic illustration of this construction is shown in Fig.~\ref{fig:ComplexRT}.

Accordingly, in $\mathrm{dS}_{d+1}^{\mathbb C}$ the Ryu--Takayanagi prescription
assigns to a subregion $A$ the pseudoentropy
\begin{equation}\label{eq:RTformula}
S_\Gamma(A)=\frac{\Area}{4\GN}\,,
\qquad
\Area=\int_{\Gamma}\diff^{d-1}\sigma\,\sqrt{\gamma}\,,
\end{equation}
evaluated on an admissible complex extremal surface $\Sigma_A$ of complex
dimension $d-1$. Here $\gamma=\det\gamma_{ab}$, with $\gamma_{ab}$ the induced
holomorphic metric, and $\Gamma\subset\Sigma_A$ is a real $(d-1)$-dimensional integration cycle with $\partial\Gamma=\partial A$, homologous to $A$.

The area integrand is defined by analytic continuation of $\sqrt{\gamma}$ from the
real Lorentzian part of $\Gamma$ adjacent to $\partial A$. There we fix its
branch such that the UV divergences are negative imaginary, as in
Ref.~\cite{Anastasiou:2026bbf}. By holomorphicity, the integral is unchanged
under continuous deformations of $\Gamma$, with $\partial\Gamma$ and the branch
there held fixed, as long as $\Gamma$ stays in the regular domain of $\Sigma_A$,
i.e.\ away from the singular points of the area integrand. The area therefore depends
on $\Gamma$ only through its relative homology class in this domain.

Extremal surfaces are determined by requiring the area to be stationary,
$\delta\operatorname{Area}_\Gamma(\Sigma_A)=0$, under variations of the
embedding with fixed boundary data and under the requirement of bulk holomorphicity. If the induced
metric is nondegenerate, the complex normal bundle admits a local orthonormal
frame $n_I^\mu$, $n_I\cdot n_J=\delta_{IJ}$. Stationarity is then equivalent to
the vanishing of the mean curvature vector,
\begin{equation}\label{eq:ExtCon}
K^I=\gamma^{ab}K^I{}_{ab}=0,\qquad I=1,2\,,
\end{equation}
where $K^I{}_{ab}$ is the extrinsic curvature along $n_I^\mu$. This condition is
independent of the choice of frame; see Appendix~\ref{app:TrK} for the
derivation.

The homology classes of $\Gamma$ relevant for a given region are organized by
discrete symmetries of the complexified geometry. The metric~\eqref{eq:dScom}
depends on $\uptau$ only through $\cosh^2\uptau$, so that the shift
\begin{equation}\label{eq:Rshift}
R:\ \uptau\longrightarrow\uptau+\iu\pi
\end{equation}
is a discrete isometry. Since it maps
$(\sinh\uptau,\cosh\uptau)\to-(\sinh\uptau,\cosh\uptau)$, it acts on
$\mathcal M_{\mathbb C}$ as the antipodal map. Its square is the identity,
because $\uptau\sim\uptau+2\pi\iu$, establishing the domain of $\uptau$ as the complex cylinder. Let $\Gamma^{(0)}$ denote the class of integration cycles
continuously connected to the HH construction, which matches it for the maximal spherical entangling region. It runs along the Lorentzian
section from $\partial A$ at $\uptau \rightarrow \infty$ and caps off smoothly at a turning point $\uptau_\star$
with $\operatorname{Im}\uptau_\star<0$; for the unit sphere it covers
$\Sigma_A^{(\mathrm t)}\cup\Sigma_A^{(\mathrm s)}$. For $n\in\mathbb Z$ we define the integration cycle
\begin{equation}\label{eq:Gamma-n}
\Gamma^{(n)}=U_n\cup R^n\big(\Gamma^{(0)}\big)\,,
\qquad
U_n:\ \tauUV\longrightarrow\tauUV+\iu\pi n\,,
\end{equation}
where $U_n$ is a path in holomorphic $\uptau$ which runs along the cutoff $\operatorname{Re}\uptau=\tauUV$, where $\tauUV \rightarrow \infty$ at the conformal boundary and $R^{n}(\Gamma^{(0)})$ denotes applying the $R$ shift $n$ times to the cycle $\Gamma^{(0)}$. Geometrically, this amounts to a translation of $\Gamma^{(0)}$ in the complex plane by $n\pi$ along a line parallel to the imaginary axis. The surface
$R^n(\Sigma_A)$ is extremal because the shift $R$ is an isometry and $\Sigma_{A}$ is already extremal. Also, the segment  $U_{n}$ renders homologous any surface $R^n(\Sigma_A)$. Then, the integration cycle  $\Gamma^{(n)}$ is anchored on the
same entangling surface, since the embedding asymptotically approaches its
boundary value uniformly in $\operatorname{Im}\uptau$. Its turning point is
$\uptau_\star+\iu\pi n$, where $n$ labels the winding around the $\uptau$-cylinder and half-winding represents the antipodal map in the spherical transverse section. To account for the different $n$-windings, which correspond to different saddles of the complex extremal surface, we define the entropy 
\begin{equation}\label{eq:Sn}
S^{(n)}(A)= S_{\Gamma^{(n)}}(A)\,,
\end{equation}
associated with the corresponding integration cycle.

As it is derived in the following sections for different cases, the standard form of the pseudoentropy expansion gets modified when accounting for the different $n$-saddles. In particular, one obtains
\begin{equation}\label{eq:Sexp}
S^{(n)}(A)=\sum_{k=1}^{\lfloor\frac{d-1}{2}\rfloor}c_{d-2k}\left(\frac{H}{\delta}\right)^{d-2k}
+\begin{cases}
(-1)^n(-1)^{\frac{d-1}{2}}\,\mathsf S_{\mathrm u}(A), & \text{odd } d\,,\\[4pt]
(-1)^{\frac{d-2}{2}}\mathsf S_{\mathrm u}(A)
\left[\log\dfrac{H}{\delta}+\iu\pi\left(\dfrac12-n\right)\right]+c_0, & \text{even } d\,,
\end{cases}
\end{equation}
where $\delta$ is the UV cutoff, $H$ is a characteristic length scale of the
entangling region, and the coefficients $c_{d-2k}$ are regularization dependent.
The term $\mathsf S_{\mathrm u}(A)$ is independent of $n$ and encodes universal data of the dual CFT. Moreover, for even $d$, an additional term $c_0$ appears, representing a scheme-dependent, non-universal finite contribution whose value can be modified by a rescaling of the  UV cutoff.

Finally, it should be emphasized that the matching conditions of the piecewise construction are
automatically satisfied by a holomorphic embedding. With
$\uptau=-\iu\tau_{\mathrm E}$, the junction condition
$\partial_\uptau=\iu\,\partial_{\tau_{\mathrm E}}$ is the chain rule, i.e.\ the
Cauchy--Riemann condition. The piecewise HH surface is therefore the restriction
of a single complex saddle to the real sections.

In what follows, we apply the holomorphic extremization procedure described above to spherical entangling surfaces in the dS geometric vacuum of arbitrary size and arbitrary dimension $d\geq3$, and small harmonic deformations thereof. The reason why we start from dS$_4$/CFT$_3$ is because the $d=2$ case has already been studied by~\cite{Fujiki:2025rtx}---see also~\cite{Narayan:2026wzp,Li:2026adq,Heller:2024whi,Heller:2025kvp,Bernamonti:2026pxo} for the related case of timelike entanglement entropy in AdS.
 
\subsection{Example: dS$_4$/CFT$_3$}\label{subsec:dS4example}

As a first application of the general formalism developed above, we consider
the holographic pseudoentropy in dS$_4$/CFT$_3$. We begin with the undeformed
spherical entangling region, for which the holomorphic extremal surface can be
obtained analytically. This example serves as a consistency check of the
holomorphic RT prescription by reproducing the known pseudoentropy obtained
from the HH piecewise construction. We then proceed to study small shape
deformations of the entangling surface, for which the holomorphic construction reproduces the piecewise result at unit radius and extends it to arbitrary radius and to the full family of turning points.
 
\subsubsection{Spherical entangling region}\label{Sectiontfo}
 
\begin{figure*}
\centering
\begin{subfigure}{.5\textwidth}
  \centering
  \includegraphics[height=0.9\linewidth]{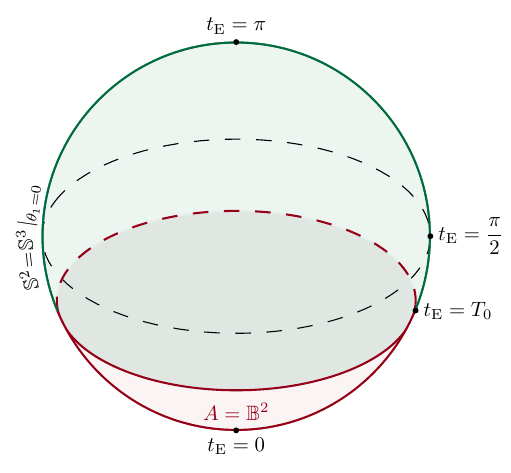}
\end{subfigure}%
\begin{subfigure}{.5\textwidth}
  \centering
  \includegraphics[height=0.9\linewidth]{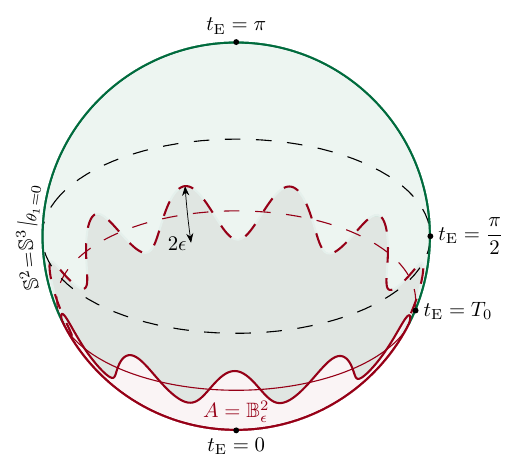}
\end{subfigure}
\caption{\justifying(Left) We show pictorially the disk entangling region $A=\mathbb B^2$ of radius $\tE=T_0$ defined in $\mathbb S^2$, which is in the equator---that is, at $\theta_1=0$ following \eqref{eq:subUndef}---of $\mathbb S^3$. (Right) We show the same disk-like entangling region of unit radius perturbed harmonically, $\mathbb B_\epsilon^2$, following \eqref{eq:phiF}.}
  \label{fig:undefdefreg}
\end{figure*}

We now evaluate the proposed holographic pseudoentropy for a spherically entangling region in the Euclidean CFT$_3$ dual to complexified global $\mathrm{dS}_{4}^{\mathbb C}$. For that we write down the complexified metric~\eqref{eq:dScom} for $\mathrm{dS}_{4}^{\mathbb C}$
\begin{equation}\label{eq:dScom4}
\diff \mathsf s^2
=
\Ls^2
\left[
-\diff\uptau^2
+\cosh^2\uptau
\left(
\diff\mathsf{t}_{\mathrm E}^2
+\sin^2\mathsf{t}_{\mathrm E}\,
(\diff\uptheta_1^2+\cos^2\uptheta_1\,\diff\upphi^2)
\right)
\right]\,,
\end{equation}
The boundary subregion $A$ is always defined on the real conformal boundary $\mathcal I_{\mathbb R}^{+}$ using the original boundary coordinates
$(\tE,\theta_i,\phi)$. Only the bulk extremal surface extending from
$\partial A$ is complexified.
Now consider a ball-shaped entangling region, $A=\mathbb B^2$ with entangling surface $\partial A=\mathbb S^1$, defined on the $\theta_1=0$ slice of the future conformal boundary $(\mathcal I_{\mathbb R}^{+}\simeq\mathbb S^3)$---see Fig.~\ref{fig:undefdefreg}),
\begin{equation} \label{eq:subUndef}
A=\left\{(\tE,\theta_1,\phi)\in\mathcal I_{\mathbb R}^{+}\;\middle|\;
0\le \tE\le T_0,\;\theta_1=0,\;0\le\phi<2\pi\right\}, \qquad 0<T_0<\pi\, .
\end{equation}
 
In the complexified bulk, we consider a complex codimension-two extremal
surface $\Sigma_A$ described by the embedding
$\mathsf{t}_{\mathrm E}=\mathsf{t}_{\mathrm E}(\uptau)$ with
$\uptheta_1=0$.\footnote{
Although the bulk angular coordinates
$\uptheta_i$ and $\upphi$ are holomorphic coordinates on
$dS_{d+1}^{\mathbb C}$, the spherical boundary subregion preserves the
rotational symmetry of the background. Since the dS isometries are
real analytic, they admit a holomorphic extension to the complexified
manifold. We therefore restrict to the symmetry-preserving complex saddle,
for which the complex angular coordinates can consistently be taken to be
constant, namely $\uptheta_1=0$ in the present case. This simplification is
specific to the undeformed spherical entangling region and will not hold
once the boundary symmetry is broken by shape deformations, where the
embedding generally acquires non-trivial angular dependence.}
The complex extremal surface $\Sigma_A$ is parameterized by the intrinsic
coordinates $(\uptau,\upphi)$, for which the induced metric $\gamma_{ab}$
takes the form
\begin{equation}\label{eq:InduceMet}
\frac{\diff \mathsf s^2_{\gamma}}{\Ls^2}=  
\left({\mathsf{t}_{\mathrm E}'}^2  \cosh^2{\uptau}-1\right)\diff\uptau^2 + \cosh^2{\uptau} \sin^2{\mathsf{t}_{\mathrm E}} \diff\upphi^2 \,,   
\end{equation}
where $\mathsf{t}_{\mathrm E}'= \partial_{\uptau} \mathsf{t}_{\mathrm E}(\uptau)$. The holomorphic area functional is therefore
\begin{equation}\label{eq:AreaUndef}
\operatorname{Area}_\Gamma(\Sigma_A)
=\Ls^2\int_{\mathcal{C}_\uptau}\diff\uptau
\int_{\mathcal{C}_\upphi}\diff\upphi\,
\cosh\uptau\,\sin\mathsf{t}_{\mathrm E}\,
\sqrt{{\mathsf{t}_{\mathrm E}'}^2\cosh^2\uptau-1}\,,
\end{equation}
for an integration cycle $\Gamma=\mathcal C_\uptau\times\mathcal C_\upphi$.
 
The Euler--Lagrange equation associated with
Eq.~\eqref{eq:AreaUndef}, or equivalently the extremality condition
$K^I=0$, admits the solution
\begin{equation} \label{eq:solUndef}
 \cos {\mathsf{t}_{\mathrm E}}(\uptau)=\cos T_0\,\tanh\uptau\,.
\end{equation}
which satisfies the required boundary condition $\big(\operatorname{Re}(\uptau)\rightarrow+\infty,
\quad
\operatorname{Im}(\uptau)=0 \big)
\Longrightarrow
\tE\rightarrow T_0,$
so that the extremal surface is anchored on the prescribed entangling
surface $\partial A$. 
Regularity of the extremal surface requires the existence of a smooth turning point $\uptau_\star$, at which the surface caps off smoothly, i.e., at which the transverse circle shrinks. On the solution~\eqref{eq:solUndef}, $\cosh^2\uptau\sin^2\mathsf t_{\mathrm E}=1+\sin^2T_0\sinh^2\uptau$, so that
\begin{equation}\label{eq:turning-eq}
1+\sin^2T_0\,\sinh^2\uptau_\star=0.
\end{equation}
For $T_0\neq\pi/2$ this is equivalent to $\diff\mathsf t_{\mathrm E}/\diff\uptau|_{\uptau_\star}=\infty$, expressing that the surface becomes tangent to the constant-$\uptau$ slice, with $\mathsf t_{\mathrm E}(\uptau_\star)\in\{0,\pi\}$. For $T_0=\pi/2$ the profile is $\mathsf t_{\mathrm E}=\pi/2$, and the surface caps off because $\cosh\uptau_\star=0$.
 
Solving Eq.~\eqref{eq:turning-eq} gives the family of turning points
\begin{equation} \label{eq:turnP}
\uptau_\star^{(n)}
= \pm \log \left(\cot\frac{T_0}{2}\right)
-\iu\pi\left(\frac12
-n\right)\,,
\qquad n\in\mathbb Z\,.
\end{equation}
The two signs correspond to $\mathsf t_{\mathrm E}\left(\uptau_\star^{(n)}\right)=0$ and $\pi$, i.e., to capping off on the side of $A$ or of its complement. Both give $\sin T_0\sinh\left(\uptau_\star^{(n)}\right)=-\iu(-1)^n$, and hence identical areas, consistent with $S(A)=S(\bar A)$ for transition matrices built from pure states. Below we take the member with $\operatorname{Re}\left(\uptau_\star^{(n)}\right)\ge0$.
The integer appearing in Eq.~\eqref{eq:turnP} admits a natural interpretation in
terms of the analytic structure of the boundary observable. A closely
related phenomenon occurs for timelike entanglement entropy in two-dimensional
CFTs~\cite{Bernamonti:2026pxo}. There, the analytic continuation of the
replica twist correlator across successive light cones moves the correlator
between different Riemann sheets. The corresponding bulk complex geodesics
extend by integer multiples of $\pi$ in the imaginary direction, precisely
mirroring the branch structure of the boundary correlator. The relevant
sheet is not arbitrary: it is fixed by the $\iu\epsilon$ prescription and,
equivalently, by the ordering of the boundary twist operators.
 
In the present dS/CFT setting, the shift $\uptau_\star^{(n)}\to\uptau_\star^{(n+1)}$
similarly originates from the multivalued inverse functions involved in
solving the turning-point condition and labels different analytic
continuations of the complex extremal surface, as discussed in Sec.~\ref{subsec:holoRT}. It is therefore natural to
interpret $n$ as a sheet label for the analytically continued
pseudoentropy.
 
Now we evaluate the area of the complex extremal surface $\Sigma_A$. Substituting the solution \eqref{eq:solUndef} into the area functional \eqref{eq:AreaUndef}, we obtain
\begin{equation}\label{eq:AreaSigmaUndef}
\operatorname{Area}_{\Gamma^{(n)}}(\Sigma_A)
=
\iu\Ls^2\sin T_0
\int_{\mathcal C^{(n)}_{\uptau}}\diff\uptau\,\cosh\uptau
\int_{\mathcal C_{\upphi}}\diff\upphi\,,
\end{equation}
where $\Gamma^{(n)}=\mathcal C^{(n)}_\uptau\times\mathcal C_\upphi$, with
$\mathcal C^{(n)}_\uptau$ the $\uptau$-contour of Eq.~\eqref{eq:Gamma-n},
oriented from the cutoff towards $\uptau_\star^{(n)}$, and $\mathcal C_\upphi$
a loop winding once around the angular cylinder.

A particular choice of $\mathcal C^{(n)}_\uptau$ integration cycle can be explicitly described by a path consisting of straight segments on the complex plane as
\begin{equation}\label{eq:contour-family}
\mathcal C^{(n)}_\uptau:\quad
\tauUV\longrightarrow\tauUV-\iu \pi \left(\frac{1}{2}- n\right)\longrightarrow\ \operatorname{Re}\uptau_\star^{(n
)}-\iu\pi \left(\frac{1}{2}- n\right)
=\uptau_\star^{(n)}\,.
\end{equation}
 
\begin{figure*}
\centering
\begin{subfigure}{.5\textwidth}
  \centering
  \includegraphics[height=0.9\linewidth]{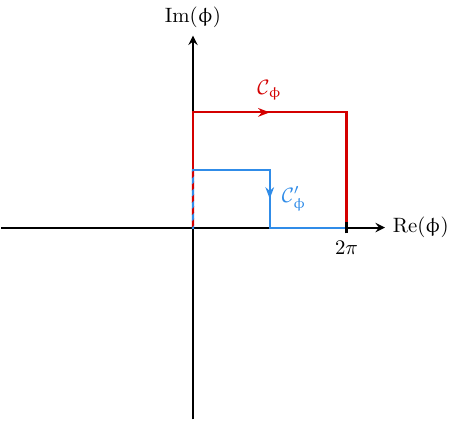}
\end{subfigure}%
\begin{subfigure}{.5\textwidth}
  \centering
  \includegraphics[height=0.9\linewidth]{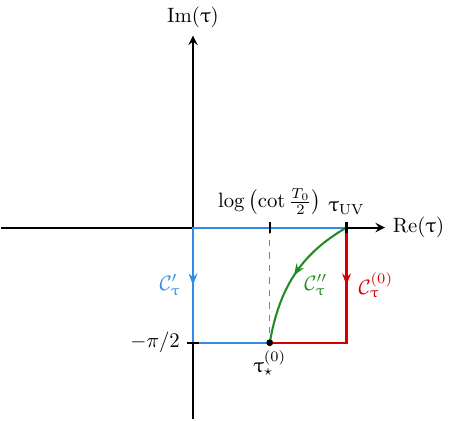}
\end{subfigure}
\caption{(Left) Representative contours $\mathcal C_\upphi\simeq\mathcal C'_\upphi$
in the complex $\upphi$-plane, connecting $\upphi=0$ to $\upphi=2\pi$; since
$\upphi\sim\upphi+2\pi$, both wind once around the angular cylinder. (Right)
Representative contours for $n=0$, from the UV endpoint $\tauUV$ to $\uptau_\star^{(0)}$: the
canonical $\mathcal C^{(0)}_\uptau$ (red), $\mathcal C'_\uptau$ through the
Lorentzian and Euclidean sections (blue), and a generic complex contour
$\mathcal C''_\uptau$ (green). Since $\cosh\uptau$ is entire, all give the same
area.}
\label{fig:contourPhiUndef}
\end{figure*}

The contour $\mathcal{C}_{\upphi}$ winds once around the complexified angular cylinder, so that
\begin{equation}
 \int_{\mathcal{C}_{\upphi}}\diff\upphi=2\pi\,.   
\end{equation}
Two representatives are shown in Fig.~\ref{fig:contourPhiUndef} (left). The
contour $\mathcal C_\upphi$ runs at constant $\operatorname{Im}\upphi>0$ over the
whole period, whereas $\mathcal C'_\upphi$ returns to the real axis before
$\operatorname{Re}\upphi=2\pi$. Both connect $\upphi=0$ to $\upphi=2\pi$. Since
$\upphi\sim\upphi+2\pi$, they are closed loops homotopic to the real circle,
$\mathcal C'_\upphi\simeq\mathcal C_\upphi$, and give the same contribution to the
area.
Furthermore, the integrand $\cosh\uptau$ is an entire function of the complex variable $\uptau$. Consequently, by Cauchy's integral theorem, the value of the integral is independent of the choice of contour connecting the endpoints $\uptau=\uptau_\star^{(n)}$ and $\operatorname{Re}(\uptau)\rightarrow+\infty$, provided no singularities are crossed during the deformation. Hence any two such contours are continuously deformable into one another and yield the same area.

The $\uptau$-integral is elementary. For any
$\mathcal C_\uptau\simeq\mathcal C^{(n)}_\uptau$, oriented from the cutoff
$\tauUV=-\log\delta$ to the turning point,
\begin{equation}\label{eq:tauIntegral}
\int_{\mathcal C_\uptau}\diff\uptau\,\cosh\uptau
=\sinh\uptau_\star^{(n)}-\sinh\tauUV
=-(-1)^n\frac{\iu}{\sin T_0}-\frac{1}{2\delta}+\mathcal O(\delta)\,,
\end{equation}
where $\sinh\left(\uptau_\star^{(n)}\right)=-\iu(-1)^n/\sin T_0$ follows from Eq.~\eqref{eq:turnP} and $\delta$ is the ultraviolet cutoff defined by $\uptau=-\log\delta$, so that
$\delta\to0$ corresponds to $\operatorname{Re}\uptau\to+\infty$.\footnote{This
cutoff is related to the Fefferman--Graham cutoff $\delta_{\rm FG}$ of
Ref.~\cite{Anastasiou:2026bbf}, defined through $\eta=2\e{-\tau}$, by
$\delta_{\rm FG}=2\delta$. In terms of $\delta_{\rm FG}$, the divergent term reads
$-\iu\pi\Ls^2\sin T_0/(2\GN\delta_{\rm FG})$, in agreement with
Ref.~\cite{Anastasiou:2026bbf} for $T_0=\pi/2$. Universal terms are unaffected.}
Since $\cosh\uptau$ is entire, the result depends only on the endpoints. This is
illustrated in Fig.~\ref{fig:contourPhiUndef} (right) for $n=0$, where the three
representatives give the same value. 

Therefore, the pseudoentropy associated with the ball-shaped subregion
$\mathbb{B}^2$ on the contour class $\Gamma^{(n)}$ is
\begin{equation}\label{eq:pseudoUndef}
S^{(n)}(\mathbb B^2)
=
\frac{\operatorname{Area}_{\Gamma^{(n)}}(\Sigma_A)}{4\GN}
=
-\frac{\iu\pi\Ls^2}{4\GN}\frac{\sin T_0}{\delta}
+\frac{(-1)^n\pi\Ls^2}{2\GN}\,,
\end{equation}
This has the form of Eq.~\eqref{eq:Sexp} for $d=3$. The universal part is
cutoff-independent and proportional to the negative-valued central charge
$a^{\star}=-\Ls^2/(4\GN)$ of the dual Euclidean CFT,
\begin{equation}\label{eq:SuB2}
\mathsf{S}_{\mathrm u}(\mathbb B^2) = -\frac{\pi\Ls^2}{2\GN} = 2\pi a^{\star}\,.
\end{equation}
For HH, i.e., $n=0$, Eq.~\eqref{eq:pseudoUndef} reproduces the piecewise
result of Ref.~\cite{Anastasiou:2026bbf}, including the positive Euclidean
contribution $2\pi\Ls^2/(4\GN)$. The holomorphic RT prescription therefore
reproduces the expected pseudoentropy for the spherical entangling region.
 
\subsubsection{Small deformation around sphere}
 
Having established that the holomorphic RT prescription reproduces the
pseudoentropy of the spherical entangling region, we now turn to a less trivial test of the proposal by considering small shape deformations of
the entangling surface. In the original HH construction, the
leading correction to the pseudoentropy was obtained from a piecewise
Lorentzian--Euclidean extremal surface. Here we revisit the same problem
within the complexified dS bulk, where the extremal surface is
described by a single holomorphic embedding. For $T_0=\pi/2$ and $n=0$, the holomorphic embedding restricted to the real and imaginary $\uptau$ axes reproduces the two segments of the piecewise surface, so agreement with Refs.~\cite{Anastasiou:2025rvz,Anastasiou:2026bbf} is expected there. The holomorphic construction extends the computation to arbitrary $T_0$, where no real piecewise surface exists, and to arbitrary $n$.
 
We now consider a small harmonic deformation $\mathbb{B}_\epsilon^2$ of the spherical entangling region about $\tE=T_0$, parameterized by
\begin{equation}\label{eq:phiF}
\mathbb{B}_\epsilon^2:\qquad
T_\epsilon
=
T_0
+
\epsilon\,\sin T_0\,\sum_{\ell}\Phi_\ell(\phi)\,,
\qquad
\Phi_\ell(\phi)
=\left[
\frac{a_\ell}{\sqrt{\pi}}\cos(\ell\phi)
+
\frac{b_\ell}{\sqrt{\pi}}\sin(\ell\phi)
\right]\,.
\end{equation}
Here $\epsilon\ll1$ is the deformation parameter, while $a_\ell$ and
$b_\ell$ specify the amplitudes of the individual Fourier modes. A
schematic illustration of the deformed entangling region is shown in
Fig.~\ref{fig:undefdefreg}.
The overall factor of $\sin T_0$ is completely fixed by the geometry of the
boundary sphere. More precisely, it follows from the stereographic map $r=\tan\frac{\tE}{2},$ which relates a radial deformation of a planar disk to a deformation of its image on $\mathbb S^2$. Expanding the stereographic map to first order in the deformation parameter yields precisely the factor $\sin T_0$ multiplying $\Phi(\phi)$. The derivation is presented in Appendix~\ref{app:stereographic}.
 
In this case, the complex codimension-two extremal surface $\Sigma_A$ is
described by the embedding
\[
\mathsf{t}_{\mathrm E}
=
\mathsf{t}_{\mathrm E}(\uptau,\upphi),
\]
which now depends explicitly on the angular coordinate $\upphi$ since the
boundary deformation breaks azimuthal symmetry. Consequently, the
induced metric on $\Sigma_A$ takes the form
\begin{equation}\label{eq:GdefLT}
\frac{\diff \mathsf s_{\gamma}^{2}}{\Ls^{2}}
=
\left(
{\mathsf{t}_{\mathrm E}'}^{2}\cosh^{2}\uptau-1
\right)\diff\uptau^{2}
+\cosh^{2}\uptau
\left(
\sin^{2}\mathsf{t}_{\mathrm E}
+\dot{\mathsf{t}}_{\mathrm E}^{\,2}
\right)\diff\upphi^{2}
+2\cosh^{2}\uptau\,
\mathsf{t}_{\mathrm E}'
\dot{\mathsf{t}}_{\mathrm E}\,
\diff\uptau\,\diff\upphi\,,
\end{equation}
where $\mathsf{t}_{\mathrm E}' = \partial_{\uptau}\mathsf{t}_{\mathrm E}(\uptau,\upphi),
\,
\dot{\mathsf{t}}_{\mathrm E} = \partial_{\upphi}\mathsf{t}_{\mathrm E}(\uptau,\upphi).$ Motivated by the boundary deformation \eqref{eq:phiF}, we consider the
following perturbative ansatz for the bulk embedding,
\begin{equation}\label{eq:tET}
\mathsf{t}_{\mathrm E}(\uptau,\upphi)
=
\arccos\!\left(
\cos T_0\,\tanh\uptau
\right)
+
\epsilon\,\sin T_0
\sum_{\ell}
\mathcal{T}_{\ell}(\uptau)
\Phi_{\ell}(\upphi)\,,
\end{equation}
where the first term is the undeformed holomorphic extremal surface found in
Eq.~\eqref{eq:solUndef}. The second term describes the linear response of the
bulk embedding to the boundary shape deformation. This separation of variables ansatz is motivated by the fact that the basis of Fourier modes is a complete basis for real functions on $\mathbb S^1$ and the bulk embedding should reproduce the entangling surface at the boundary. It is the natural
holomorphic generalization of the ansatz employed in
the HH construction in our previous works~\cite{Anastasiou:2025rvz,Anastasiou:2026bbf}, with the
difference that both the embedding coordinates and the radial profile
functions are now defined on the complexified bulk geometry. The radial
functions $\mathcal{T}_{\ell}(\uptau)$ are required to satisfy the appropriate
real boundary conditions so that the complex extremal surface is anchored on
the prescribed deformation of the real conformal boundary.

Substituting the ansatz \eqref{eq:tET} into the extremality condition
\eqref{eq:ExtCon} and expanding to linear order in $\epsilon$, we obtain
\begin{equation}\label{eq:diffeq1}
\mathcal T_\ell''
+\left[
\frac{4\sin^2T_0\sinh\uptau\cosh\uptau}
{1+\sin^2T_0\sinh^2\uptau}
-\tanh\uptau
\right]\mathcal T_\ell'
+\frac{(\ell^2-1)\sin^2T_0\cosh^2\uptau}
{\left(1+\sin^2T_0\sinh^2\uptau\right)^2}
\mathcal T_\ell=0\,,
\end{equation}
where primes denote derivatives with respect to $\uptau$.
We impose the boundary conditions

\begin{equation}\label{eq:BC4D}
\lim_{\substack{\operatorname{Re}\uptau\to+\infty\\
\operatorname{Im}\uptau=0}}
\mathcal T_\ell(\uptau)=1,
\qquad
\mathcal T_\ell\ \text{regular at}\ \uptau_\star^{(n)}\,.
\end{equation}

The first condition reproduces the prescribed boundary deformation
\eqref{eq:phiF}.
Near a generic turning point $\uptau_\star^{(n)}$, the two independent solutions behave as
$\left(1+\sin^2T_0\sinh^2\uptau\right)^{(\ell-1)/2}$ and
$\left(1+\sin^2T_0\sinh^2\uptau\right)^{-(\ell+1)/2}$.
Regularity selects the first behavior, so
$\mathcal T_\ell\left(\uptau_\star^{(n)}\right)=0$ for $\ell\ge2$. The regularity requirement is physically motivated as the turning point is part of the complex RT surface, and therefore it lies inside the domain of the area functional which should remain finite in the bulk. This condition ensures that the location of $\uptau_\star^{(n)}$ is the same as for the undeformed case. Furthermore,
the maximal entangling region case is recovered by taking the continuous limit
$T_0\to\pi/2$.

For the turning-point family satisfying
$\sin T_0\sinh\left(\uptau_\star^{(n)}\right)=-\iu(-1)^n$, the normalized regular
solution is
\begin{equation}\label{eq:taucomplex}
\mathcal T_\ell(\uptau)
=
\frac{
\left(\sin T_0\sinh\uptau+\iu(-1)^n\right)^\ell
\left(\sin T_0\sinh\uptau-\iu(-1)^n\ell\right)
}{
\left(1+\sin^2T_0\sinh^2\uptau\right)^{(\ell+1)/2}
}\,.
\end{equation}
At the real UV endpoint, the denominator is chosen positive.
Its value elsewhere is obtained by analytic continuation along
the chosen contour, without resetting it to its principal value.
For $\ell=1$, the solution reduces to $\mathcal T_1=1$, which does not satisfy the required condition of $\mathcal T_1\left(\uptau_\star^{(n)}\right)=0$. However, as we are interested in proper shape deformations, we restrict to $\ell\ge2$ in what follows.\footnote{As usual, $\ell=0$ corresponds to a dilatation and $\ell=1$ to a translation of the spherical entangling surface, and not to proper deformations thereof.}

Substituting Eq.~\eqref{eq:taucomplex} into the ansatz
\eqref{eq:tET} determines the embedding to first order in
$\epsilon$. We then use the induced metric \eqref{eq:GdefLT}
to extract the universal quadratic correction to the area. The local behavior at the turning points and the lack of singularities
of the area integrand at order $\epsilon^2$---required for contour independence of the area through Cauchy's theorem---are discussed in
Appendix~\ref{app:turningpoint}.

Extracting the universal contribution, we obtain
\begin{align}\label{eq:deform4D}
\mathsf{S}_{\mathrm u}(\mathbb{B}_\epsilon^{2})
=
- \frac{\pi\Ls^2}{2\GN}
-\frac{\Ls^2}{8\GN}\,
\epsilon^2
\sum_{\ell}
\ell(\ell^2-1)
\left(a_\ell^2+b_\ell^2\right),
\end{align}
independently of $T_0$. This reproduces the HH result
\cite{Anastasiou:2025rvz,Anastasiou:2026bbf},
providing a consistency check of the holomorphic prescription
beyond spherical symmetry. The dependence on $n$ is given
in Eq.~\eqref{eq:Sexp}.
 
Following the analogy with the AdS/CFT correspondence, we introduce the quantity
\begin{equation}\label{eq:CT3d}
C_T^{\mathrm{dS}}
=
-\frac{3\Ls^2}{\pi^3\GN}\,,
\end{equation}
which was conjectured in Refs.~\cite{Anastasiou:2025rvz,Anastasiou:2026bbf} to
play the role of the coefficient controlling the CFT stress tensor two-point function \begin{equation} \langle T_{\alpha\beta}(x)T_{\gamma\delta}(0)\rangle=C_T/x^{2d}\left(I_{\alpha(\gamma}I_{\delta)\beta}-\delta_{\alpha\beta}\delta_{\gamma\delta}/d\right)\,,
\end{equation} 
where $I_{\alpha\beta}=\delta_{\alpha\beta}-2x_\alpha x_\beta/x^2$ in general dimension~\cite{Osborn:1993cr}. In terms of Eqs.~\eqref{eq:SuB2} and~\eqref{eq:CT3d}, Eq.~\eqref{eq:deform4D} reads
\begin{align}\label{eq:FAdef4D}
\mathsf{S}_{\mathrm u}(\mathbb B_\epsilon^{2})
=
2\pi a^\star
+
\frac{\pi^3C_T^{\mathrm{dS}}}{24}
\epsilon^2
\sum_\ell
\ell(\ell^2-1)
(a_\ell^2+b_\ell^2)\,.
\end{align}
This corresponds precisely to the conjectured Mezei-like formula for holographic pseudoentropy in dS$_4$/CFT$_3$~\cite{Allais:2014ata,Mezei:2014zla,Faulkner:2015csl}. 
Like its AdS counterpart, the leading correction shares the same sign as the finite piece. In AdS, where $a^\star>0$ and $C_T>0$, the spherical entangling surface is therefore a local minimum of $\mathsf S_{\rm u}$, i.e., a local maximum of the finite part of the entanglement entropy. In dS$_4$, both $a^\star$ and $C_T^{\rm dS}$ are negative and the sphere is a local maximum of $\mathsf S_{\rm u}$. This is a direct consequence of the analytic continuation $\left.L\right|_{\text{AdS}}\rightarrow-\iu \left.L\right|_{\text{dS}}$~\cite{Anastasiou:2025rvz,Anastasiou:2026bbf}.\footnote{More generally, entanglement entropy for arbitrary unitary three dimensional CFT is globally maximized by the spherical entangling surface~\cite{Fonda:2015nma,Anastasiou:2020smm,Anastasiou:2021swo,Bueno:2021fxb,Anastasiou:2022pzm,Bueno:2023gey}, whether a global bound also holds for pseudoentropy---even in the holographic case---remains unclear. For holographic entanglement entropy in this dimension, global maximization follows from its relation to the Willmore energy~\cite{willmore1965note,willmore1996riemannian,marques2014min,Toda2017Willmore,Babich:1992mc,alexakis2010renormalized}, a relation that has no counterpart in dS space. See also~\cite{Astaneh:2014uba,Perlmutter:2015vma} for more results on shape dependence employing Willmore energy for CFTs in four dimensions and~\cite{guven2005conformally,Graham:2017bew,Zhang:2017lcd,Martino:2023oog,Anastasiou:2024rxe,Lan:2025evd,Bueno:2026hku} for five dimensions.} 
 
In conclusion, we observe that our conjectured relation for $C_T^{\mathrm{dS}}$, \eqref{eq:FAdef4D}, first proposed employing the HH prescription in~\cite{Anastasiou:2025rvz,Anastasiou:2026bbf} continues to hold unchanged under the holomorphic RT prescription introduced in Sec.~\ref{sec:hes}.

\section{Higher dimensions}\label{sec:hdim}
\subsection{Spherical entangling regions}
\label{subsec:higherdim_sphere}
 
We now generalize the spherical example of Sec.~\ref{subsec:dS4example} to
arbitrary $\mathrm{dS}_{d+1}^{\mathbb C}/\mathrm{CFT}_d$, and show that the
holomorphic RT prescription can be solved in closed form for any $d$,
with the 4D result of Sec.~\ref{sec:hes} arising as the special
case $d=3$.
We work with the complexified metric in Eq.~\eqref{eq:dScom}.
We consider the ball-shaped entangling region $A=\mathbb B^{d-1}$ on the
real future conformal boundary $\mathcal I_{\mathbb R}^+\simeq\mathbb S^d$,
defined on the equatorial slice $\theta_1=0$,
\begin{equation}\label{eq:subHigherD}
    A=\Big\{(t_{\mathrm E},\theta_1,\theta_2,\ldots,\theta_{d-2},\phi)
    \in\mathcal I_{\mathbb R}^+\;\Big|\;
    \theta_1=0,\;\;0\le t_{\mathrm E}\le T_0\Big\},
    \qquad 0<T_0<\pi\,,
\end{equation}
with entangling surface $\partial A=\{t_{\mathrm E}=T_0,\theta_1=0\}\cong\mathbb S^{d-2}$. As in the $\mathrm{dS}_4/\mathrm{CFT}_3$ case, rotational invariance allows
us to restrict to the symmetry-preserving complex saddle
$\mathsf t_{\mathrm E}=\mathsf t_{\mathrm E}(\uptau)$, $\uptheta_1=0$, with
the surface spanning the remaining $\mathbb S^{d-2}$. Since
$\diff\Omega_{d-1}^2|_{\theta_1=0}=\diff\Omega_{d-2}^2$, the induced
holomorphic metric on $\Sigma_A$ is
\begin{equation}\label{eq:InduceMetD}
    \frac{\diff\mathsf s_\gamma^2}{\Ls^2}
    =\Big({\mathsf t_{\mathrm E}'}^2\cosh^2\uptau-1\Big)\diff\uptau^2
    +\cosh^2\uptau\,\sin^2\mathsf t_{\mathrm E}\;\diff\Upomega_{d-2}^2\,,
\end{equation}
which generalizes Eq.~\eqref{eq:InduceMet}. Writing the integration cycle as
$\Gamma=\mathcal C_\uptau\times\mathbb S^{d-2}$, the holomorphic area functional is
\begin{equation}\label{eq:AreaD}
    \operatorname{Area}_{\Gamma}(\Sigma_A)
    =\Ls^{d-1}\,\mathrm{Vol}(\mathbb S^{d-2})
    \int_{\mathcal C_\uptau}\diff\uptau\;
    \cosh^{d-2}\uptau\,\sin^{d-2}\mathsf t_{\mathrm E}\,
    \sqrt{{\mathsf t_{\mathrm E}'}^2\cosh^2\uptau-1}\, ,
\end{equation}
where 
\begin{equation}\label{eq:VolS}
    \operatorname{Vol}(\mathbb S^{d-2})=\frac{2\pi^{(d-1)/2}}{\Gamma[(d-1)/2]}
\end{equation}
denotes the volume of the transverse sphere. In the complexified bulk, this volume contribution is the same as for a real spherical transverse section because the area integrand is independent of the angular variables and thus it is the integration of a trivial entire function subject to the real boundary condition. For every $d$, the extremality condition $K^I=0$
is solved by the profile in Eq.~\eqref{eq:solUndef}, as can be verified
by substitution into the Euler--Lagrange equation derived from
Eq.~\eqref{eq:AreaD}. This profile satisfies the same boundary condition
and turning-point regularity condition, Eq.~\eqref{eq:turning-eq},
so its turning points are again given by Eq.~\eqref{eq:turnP}.
The profile and turning-point structure are therefore independent of
$d$; all dimensional dependence of the pseudoentropy resides in the
area integral~\eqref{eq:AreaD}.

Substituting Eq.~\eqref{eq:solUndef} into Eq.~\eqref{eq:AreaD}, the
area functional takes the form
\begin{equation}\label{eq:AreaMaster}
\operatorname{Area}_{\Gamma}(\Sigma_A)
=
\iu\,\Ls^{d-1}\sin T_0\,
\operatorname{Vol}(\mathbb S^{d-2})
\int_{\mathcal C_\uptau}\diff\uptau\,
\cosh\uptau\,
\Big(1+\sin^2T_0\sinh^2\uptau\Big)^{\frac{d-3}{2}}\,,
\end{equation}
which reduces to Eq.~\eqref{eq:AreaSigmaUndef} for $d=3$.
 
For odd $d$, the exponent $(d-3)/2$ is an integer, so that the
integrand in Eq.~\eqref{eq:AreaMaster} is an entire function of
$\uptau$. Consequently, for fixed endpoints, the contour
$\mathcal C_\uptau$ can be continuously deformed without encountering
singularities, and the area depends only on its endpoints.
 
For even $d$, the exponent $(d-3)/2$ is a half-integer. The
turning points determined by Eq.~\eqref{eq:turning-eq} then become
branch points of the area integrand. The
area integrand is therefore naturally viewed as a differential on the
corresponding branched Riemann surface rather than as a single-valued
function on the complex $\uptau$-plane.
 
\begin{figure*}
\centering
\begin{subfigure}{.5\textwidth}
  \centering
  \includegraphics[height=0.9\linewidth]{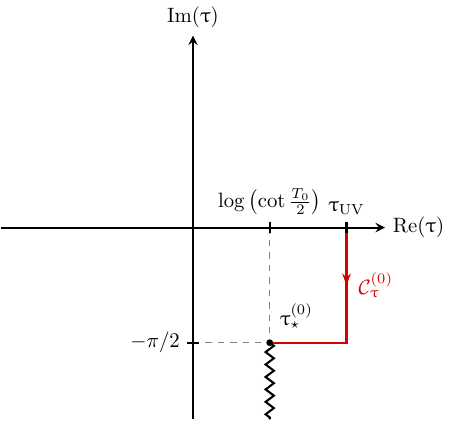}
\end{subfigure}%
\begin{subfigure}{.5\textwidth}
  \centering
  \includegraphics[height=0.9\linewidth]{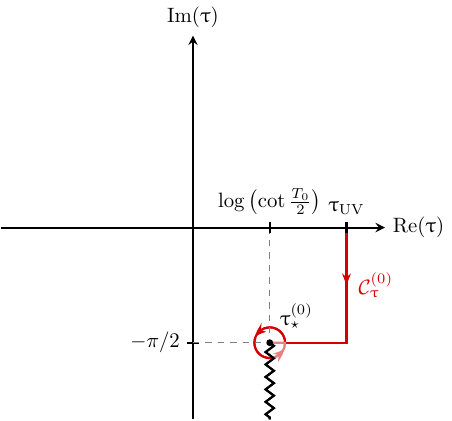}
\end{subfigure}
\caption{
Contours from $\uptau_{\mathrm{UV}}$ to $\uptau_\star^{(0)}$ for even $d$.
(Left) Direct approach to the turning point.
(Right) One winding around the turning point followed by an
approach on the second sheet, shown in a lighter shade.
The wavy lines denote branch cuts. Both contours belong
to the $n=0$ class and give the same area.
}
\label{fig:turning-point-contour}
\end{figure*}

A branch cut may be introduced to represent a particular choice of
sheet in the $\uptau$-plane, but its precise location is conventional.
The invariant information is instead contained in the ramification
points and in the analytic continuation between the different sheets
of the Riemann surface. In particular, a contour winding around a
turning point in the $\uptau$-plane changes sheets under analytic
continuation. Fig.~\ref{fig:turning-point-contour} illustrates this local
continuation for an even-dimensional boundary.

For contours terminating at the same turning point, however, this
sheet change does not lead to an additional contribution to the area.
To see this locally, one may introduce a uniformizing coordinate
\begin{equation}
w^2=\uptau-\uptau_\star^{(n)}\,.
\end{equation}
The turning point is then mapped to the regular point $w=0$ on the
Riemann surface, where the area differential is regular. Thus, a contour that winds around the turning point and subsequently
reaches the same ramification point on the second sheet is equivalent,
as a path on the Riemann surface, to the direct contour terminating at
the turning point. We illustrate this explicitly for $d=4$ in
Appendix~\ref{app:branch_contour}. Note, however, that this is a local statement. By the same reasoning discussed in Sec.~\ref{Sectiontfo}, contours that differ globally by encircling other branch points or by winding around the $\uptau$-cylinder at large $\operatorname{Re}\uptau$ can pick up nonzero monodromies, as shown in Appendix~\ref{subsec:nshift}.

To make the dependence on $n$ explicit, we use the family of contours defined in Eq.~\eqref{eq:contour-family}, made of straight segments, with $\tauUV=-\log\delta$ and $\uptau_\star^{(n)}$ as given by Eq.~\eqref{eq:turnP} with $\operatorname{Re}\uptau_\star\ge0$.
Evaluating Eq.~\eqref{eq:AreaMaster} with the cutoff
$\tauUV=-\log\delta$, we obtain the expansion in
Eq.~\eqref{eq:Sexp}, with $H=\sin T_0$ and
\begin{align}\label{eq:divergent-coefficients}
c_{d-2k}
&=
-\frac{\iu\Ls^{d-1}\pi^{d/2-1}
\Gamma\!\left(d/2-k\right)}
{4\GN\,\Gamma(d-2k+1)\Gamma(k)}
\,{}_2F_1\!\left(
1-k,\frac d2-k;d-2k+1;\sin^2 T_0
\right)\,,\\
c_0&=
-\frac{\iu\Ls^{d-1}\pi^{d/2-1}}
{4\GN\,\Gamma(d/2)}
H_{d/2-1}^{(1)}\!\left(\cos^2 T_0\right)\,,\qquad \text{for }d=\text{even}\,,
\end{align}
with $k=1,\ldots,\left\lfloor(d-1)/2\right\rfloor$ and $H_m^{(1)}(z)=\sum_{j=1}^{m}z^j/j$.
These coefficients depend on the cutoff prescription and on $H$,
but are independent of $n$. The universal coefficient is
\begin{equation}\label{eq:Sdef1}
\mathsf S_{\mathrm u}(\mathbb B^{d-1})
=
\begin{cases}
\displaystyle
\frac{(-\iu)^{d-1}\pi^{d/2}\Ls^{d-1}}
{4\GN\Gamma\left(d/2\right)}
=2\pi a^\star\,, & d\ \mathrm{odd}\,,\\[12pt]
\displaystyle
\frac{(-\iu)^{d-1}\pi^{(d-2)/2}\Ls^{d-1}}
{2\GN\,\Gamma\left(d/2\right)}
=4a^\star\,, & d\ \mathrm{even}\,,
\end{cases}
\end{equation}
where we defined
\begin{equation}
a^\star
=
\frac{(-\iu)^{d-1}\pi^{(d-2)/2}\Ls^{d-1}}
{8\,\Gamma\left(d/2\right)\GN}\,.
\end{equation}
This quantity agrees with the complex-valued central charge of the
nonunitary $\mathrm{CFT}_d$ obtained in dS/CFT
\cite{Maldacena:2002vr}. Here we display only
$\mathsf S_{\mathrm u}(\mathbb B^{d-1})$, which is independent of both
$T_0$ and $n$. The signs, the dependence on $n$, and the logarithmic
term for even $d$ follow from Eq.~\eqref{eq:Sexp}, with $H=\sin T_0$.
For $d=3$, this gives
$\mathsf S_{\mathrm u}(\mathbb B^2)=-\pi\Ls^2/(2\GN)$ and reproduces
Eq.~\eqref{eq:pseudoUndef} upon substitution into Eq.~\eqref{eq:Sexp}. The conventions of
Refs.~\cite{Anastasiou:2025rvz,Anastasiou:2026bbf}
correspond to $n=0$ and $T_{0}=\frac{\pi}{2}$ for both odd and even $d$.
  
\subsection{Small deformation around sphere}
 
We now generalize the shape-deformation analysis to arbitrary boundary
dimension $d$. We consider a small deformation of the spherical entangling
surface around $\mathsf{t}_{\mathrm E}=T_0$, with
$0<T_0<\pi$. The deformed entangling surface is defined on the real
conformal boundary $\mathcal I_{\mathbb R}^{+}\simeq\mathbb S^d$ and is
parameterized by
\begin{equation}\label{eq:defHigherD}
    \mathbb B_{\epsilon}^{d-1}:
    \
    \mathsf{t}_{\mathrm E}
    =
    T_0
    +\epsilon\,\sin T_0
    \sum_{\ell,\mathbf m}
    a_{\ell,\mathbf m}\,
    Y_{\ell,\mathbf m}(\Omega_{d-2})\,,
    \qquad
    \epsilon\ll1\,,
\end{equation}
where $Y_{\ell,\mathbf m}=Y_{\ell,\mathbf m}(\Omega_{d-2})$ are real scalar
spherical harmonics on the entangling sphere $\mathbb S^{d-2}$ with quantum numbers $\ell$, $\mathbf{m}=\{m_1,\ldots m_{d-3}\}$, orthonormal with respect to the unit-sphere measure. The spherical harmonics satisfy
\begin{equation}
    \Delta_{\mathbb S^{d-2}}Y_{\ell,\mathbf m}
    =
    -\ell(\ell+d-3)Y_{\ell,\mathbf m}\,,
\end{equation}
where $\Delta_{\mathbb S^{d-2}}$ is the Laplace--Beltrami operator on
$\mathbb S^{d-2}$, and $a_{\ell,\mathbf m}$ are the corresponding
deformation amplitudes. The factor $\sin T_0$ arises naturally from the
stereographic map as we have already discussed; the details are given in Appendix~\ref{app:stereographic}.
 
The undeformed spherical extremal surface is described by \eqref{eq:solUndef}
which is independent of the boundary dimension $d$. We therefore consider
the perturbative bulk embedding
\begin{equation}\label{eq:tETHigherD}
    \mathsf{t}_{\mathrm E}(\uptau,\Upomega_{d-2})
    =
    \arccos\!\left(
\cos T_0\,\tanh\uptau
\right)
    +
    \epsilon\,\sin T_0
    \sum_{\ell,\mathbf m}
    \mathcal T_\ell(\uptau)\,
    a_{\ell,\mathbf m}\,
    Y_{\ell,\mathbf m}(\Upomega_{d-2})
    +\mathcal O(\epsilon^2)\,.
\end{equation}
The angular dependence of the bulk extremal surface is therefore inherited
from the boundary deformation through the corresponding scalar hyperspherical
harmonics. As in the case of $d=3$, the ansatz is motivated by the completeness and orthonormality of the basis at the real conformal boundary. Its extension into the bulk through the embedding function is required by the matching of the RT surface with the entangling region as the boundary condition of the extremization problem.
 
Substituting Eq.~\eqref{eq:tETHigherD} into the holomorphic extremality
condition~\eqref{eq:ExtCon} and expanding to first order in $\epsilon$, the
equations of motion separate into independent equations for each harmonic
mode.
The resulting radial equation takes the form
\begin{equation}\label{eq:diffeqHigherD}
\mathcal T_\ell''
+\left[
\frac{(d+1)\sin^2T_0\sinh\uptau\cosh\uptau}
{1+\sin^2T_0\sinh^2\uptau}
-\tanh\uptau
\right]\mathcal T_\ell'
+\frac{(\ell-1)(\ell+d-2)\sin^2T_0\cosh^2\uptau}
{\left(1+\sin^2T_0\sinh^2\uptau\right)^2}
\mathcal T_\ell=0\,.
\end{equation}
We impose the same UV normalization and turning-point regularity
conditions as in Eq.~\eqref{eq:BC4D}, with $\uptau_\star^{(n)}$ given by
Eq.~\eqref{eq:turnP}. Near a generic turning point, the regularity of the solution
selects the exponent $(\ell-1)/2$ over $(2-\ell-d)/2$, so that
$\mathcal T_\ell\left(\uptau_\star^{(n)}\right)=0$ for $\ell\ge2$.
The local analysis is presented in Appendix~\ref{app:turningpoint}.

The resulting normalized solution is
\begin{align}\label{eq:taucomplexGeneral}
\mathcal T_\ell(\uptau)
={}&
\frac{
[\iu(-1)^n]^{\ell-1}
\Gamma(\ell+d-1)\,
\Gamma\!\left[(d+1)/2\right]
}{
2^{\ell-1}\Gamma(d)\,
\Gamma\!\left[\ell+(d-1)/2\right]
}
\left(1+\sin^2T_0\sinh^2\uptau\right)^{\frac{\ell-1}{2}}
\nonumber\\
&\times
{}_2F_1\!\left(
\ell-1,\ell+d-1;
\ell+\frac{d-1}{2};
\frac{1-\iu(-1)^n\sin T_0\sinh\uptau}{2}
\right).
\end{align}

The functions are fixed at the real UV endpoint and analytically
continued along the chosen contour, as in Eq.~\eqref{eq:taucomplex}. For $\ell=1$, $\mathcal T_1=1$ and therefore the boundary condition at $\uptau_\star^{(n)}$ is not satisfied, which is not a problem as the physically relevant deformations only consider $\ell \geq 2$. Eq.~\eqref{eq:taucomplexGeneral} reduces to
Eq.~\eqref{eq:taucomplex} for $d=3$. At $n=0$ and $T_0=\pi/2$,
its real-$\uptau$ restriction and continuation to
$\uptau=-\iu\tau_{\mathrm E}$ recover the corresponding HH
profiles~\cite{Anastasiou:2025rvz,Anastasiou:2026bbf}.

Substituting Eq.~\eqref{eq:taucomplexGeneral} into the embedding
\eqref{eq:tETHigherD}, we extract the universal contribution to the
pseudoentropy. Since the sphere extremizes this contribution, the
linear term in $\epsilon$ vanishes, and the leading shape correction
is quadratic. Introducing
\begin{equation}\label{eq:CTgeneral}
C_T^{\rm dS}
=
\frac{(-\iu)^{d-1}\Gamma(d+2)}
{8\pi^{(d+2)/2}(d-1)\Gamma(d/2)}
\frac{\Ls^{d-1}}{\GN}\,,
\end{equation}
we obtain the leading correction
\begin{align}\label{eq:MezeiGeneral}
\mathsf S_{\mathrm u}\big(\mathbb B_\epsilon^{d-1}\big)
={}&
\mathsf S_{\mathrm u}\big(\mathbb B^{d-1}\big)
+\epsilon^2 C_T^{\rm dS}
\frac{\pi^{(d+2)/2}(d-1)}
{2^{d-2}\Gamma(d+2)\Gamma(d/2)}
\sum_{\ell\ge2,\mathbf m}
a_{\ell,\mathbf m}^2\,(\ell-1)_d
\begin{cases}
\pi/2, & d\ \text{odd},\\
1, & d\ \text{even},
\end{cases}
\end{align}
where $(x)_k=\Gamma(x+k)/\Gamma(x)$ is the Pochhammer symbol
and the spherical value is given in Eq.~\eqref{eq:Sdef1}.
The result is independent of $T_0$ with the deformation
normalization given in Eq.~\eqref{eq:defHigherD}, and agrees with the
Mezei-like formula of Ref.~\cite{Anastasiou:2025rvz}.
Here we display only $\mathsf S_{\mathrm u}$; the signs and
the dependence on $n$ in the pseudoentropy follow
Eq.~\eqref{eq:Sexp}. For $d=3$, Eqs.~\eqref{eq:CTgeneral}
and \eqref{eq:MezeiGeneral} reduce to Eqs.~\eqref{eq:CT3d}
and \eqref{eq:FAdef4D}, respectively.

\section{Excited states: conical defects in dS$_3$}\label{sec:ds3-conical-defect}

Having applied the holomorphic extremization method to shape deformations of entangling regions for the geometric bulk vacuum, we now study excitations around the vacuum. This is important in order to assess whether or not the contour homology classes are maintained or whether the degeneracy on integration cycles is lifted in the excited case. For testing the method, we study the first law of pseudoentropy, which requires both the change in pseudoentropy and in modular Hamiltonian with respect to the vacuum. Therefore, we restrict to dS$_3$/CFT$_2$, as in this case, for the family of excitations considered, the energy density is known from the CFT side (see, e.g., Refs.~\cite{Hikida:2022ltr,Asplund:2014coa}).

The class of excited states here considered have as their geometric dual the three-dimensional Schwarzschild--dS family. These spacetimes
are locally dS away from conical defects and possess a cosmological horizon,
but no separate black-hole event horizon~\cite{Myung:2001ab,DeserJackiw1984,Balasubramanian2002}. Their global departure from pure dS
provides a complementary example in which the extremal embedding can still
be obtained analytically. We compute the pseudoentropy of an interval,
compare it with the vacuum at fixed boundary data, and examine the first law
in the limit of a small defect.

The metric takes the form
\begin{equation}\label{eq:cd-metric}
 \diff s^2
 =-f(r)\diff t^2+\frac{\diff r^2}{f(r)}+r^2\diff\phi^2,
 \qquad
 f(r)=\alpha^2-\frac{r^2}{\Ls^2},
 \qquad
 \phi\sim\phi+2\pi\,,
\end{equation}
where $0<\alpha\leq1$ parametrizes the conical deficit angle $2\pi(1-\alpha)$,
with pure dS recovered at $\alpha=1$. The cosmological horizon lies at
$r=\alpha\Ls$ and has inverse temperature $2\pi\Ls/\alpha$.
In the exterior region $r>\alpha\Ls$, the transformation
$r=\alpha\Ls\cosh\tau\sin\tE$ and
$\tanh(\alpha t/\Ls)=\coth\tau\cos\tE$ brings the metric to the global form
\begin{equation}\label{eq:cd-metric-global}
 \diff s^2
 =\Ls^2\left[
 -\diff\tau^2+\cosh^2\tau
 \left(\diff\tE^2+\alpha^2\sin^2\tE\,\diff\phi^2\right)
 \right]\,.
\end{equation}
For $d=2$ and $\alpha=1$, this reduces to Eq.~\eqref{eq:GdS1}.
For $\alpha<1$, the defect worldlines are located at $\tE=0,\pi$.
We formulate the extremization problem in the complexified manifold $\mathcal{M}_{\mathbb{C}}$ by
promoting $\tau,\tE,\phi\longrightarrow
\uptau,\mathsf t_{\mathrm E},\upphi\in\mathbb C$, with
$\upphi\sim\upphi+2\pi$, and taking the holomorphic extension of
Eq.~\eqref{eq:cd-metric-global}. The boundary endpoints are kept on the real
section at future infinity.

To interpret the defect contribution as a change of state, the comparison
with the vacuum must keep the boundary metric, interval size and near-boundary cutoff
fixed. Since the global metric representative
$\Ls^2(\diff\tE^2+\alpha^2\sin^2\tE\,\diff\phi^2)$ depends on $\alpha$,
we introduce a real boundary polar coordinate $\vartheta$ and choose a
common round representative between the conically-singular and vacuum configurations through
\begin{equation}\label{eq:cd-weyl}
 \tan\frac{\tE}{2}
 =\left(\tan\frac{\vartheta}{2}\right)^\alpha,
 \qquad
 \frac{\Ls^2\sin^2\vartheta}{\alpha^2\sin^2\tE}
 \left(\diff\tE^2+\alpha^2\sin^2\tE\,\diff\phi^2\right)
 =
 \Ls^2\left(\diff\vartheta^2+\sin^2\vartheta\,\diff\phi^2\right) \,.
\end{equation}
This transformation is smooth away from the defect insertions at the
poles, and at $\alpha=1$ it reduces to $\vartheta=\tE$. We hold the round representative and its dimensionless near-boundary cutoff
$\delta$ fixed, implementing the latter in global coordinates as
\begin{equation}\label{eq:cd-cutoff}
 \e{-\tau}\frac{\sin\vartheta}{\alpha\sin\tE}=\delta \,.
\end{equation}
We consider an interval $A$ on the equator
$\vartheta=\tE=\pi/2$, with real endpoints $\phi=\pm\phi_0/2$ and
$0<\phi_0<\pi$. Let $\tau_{\rm c}$ denote the
real global time at which the equatorial endpoints meet the cutoff
surface, so that the regulated geodesic is anchored at
$(\uptau,\mathsf t_{\mathrm E},\upphi)
=(\tau_{\rm c},\pi/2,\pm\phi_0/2)$.
Equation~\eqref{eq:cd-cutoff} then gives
$\tau_{\rm c}=-\log(\alpha\delta)$, which approaches future infinity
as $\delta\to0$ at fixed $\alpha$, and hence
\begin{equation}\label{eq:cd-global-cutoff}
 \e{-\tau_{\rm c}}=\alpha\delta,
 \qquad
 \cosh\tau_{\rm c}
 =\frac{1+\alpha^2\delta^2}{2\alpha\delta},
 \qquad
 \coth\tau_{\rm c}
 =\frac{1+\alpha^2\delta^2}{1-\alpha^2\delta^2}\,.
\end{equation}
In the vacuum, this prescription becomes $\e{-\tau_{\rm c}}=\delta$.\footnote{Note that for regulating the geodesic, one holds the $\delta$ regulator of the Fefferman--Graham expansion fixed. Holding $\tau_{\rm c}$ fixed instead as $\alpha$ varies would keep the
cutoff fixed in the conical representative and would therefore fail to
implement the comparison on a common round background necessary to compute the difference in geodesic length with respect to the vacuum.}

In three bulk dimensions, the holomorphic extremality condition
\eqref{eq:ExtCon} reduces to the geodesic equation. The regulated embedding is
\begin{equation}\label{eq:cd-global-geodesic}
 \mathsf t_{\mathrm E}=\frac{\pi}{2},
 \qquad
 \cos(\alpha\upphi)
 =
 \frac{1+\alpha^2\delta^2}{1-\alpha^2\delta^2}
 \cos\frac{\alpha\phi_0}{2}\,\tanh\uptau\,.
\end{equation}
We take the cutoff sufficiently small such that
\begin{equation}\label{eq:cd-kappa}
 0<
 \frac{1+\alpha^2\delta^2}{1-\alpha^2\delta^2}
 \cos\frac{\alpha\phi_0}{2}
 <1\,.
\end{equation}
Extremality can be verified using the coordinates of the embedding space
$X^0=\Ls\sinh\uptau$,
$X^1=\Ls\cosh\uptau\cos\mathsf t_{\mathrm E}$,
$X^2=\Ls\cosh\uptau\sin\mathsf t_{\mathrm E}\cos(\alpha\upphi)$
and $X^3=\Ls\cosh\uptau\sin\mathsf t_{\mathrm E}\sin(\alpha\upphi)$.
In these coordinates, Eq.~\eqref{eq:cd-global-geodesic} describes the
intersection of the dS quadric with the plane
\begin{equation}
 X^1=0,
 \qquad
 X^2=
 \frac{1+\alpha^2\delta^2}{1-\alpha^2\delta^2}
 \cos\left(\frac{\alpha\phi_0}{2}\right)X^0\,,
\end{equation}
and is therefore a geodesic.\footnote{The argument relies on the well-known fact that the intersection of a sphere with a plane that passes through its center is a maximal circle on the sphere and therefore a geodesic. Then, by analytic continuation, the sphere gets mapped to the dS quadric---whose real section is the hyperboloid---and the corresponding intersection with the plane is the geodesic here considered.} Upon removing the cutoff, the profile becomes
\begin{equation}
 \cos(\alpha\upphi)
 =\cos\frac{\alpha\phi_0}{2}\,\tanh\uptau\,.
\end{equation}
This has the form of Eq.~\eqref{eq:solUndef}, with
$T_0=\alpha\phi_0/2$ in the local angular coordinate. For $\alpha=1$,
a rotation of the round boundary identifies this embedding with the vacuum
spherical interval of radius $T_0=\phi_0/2$.

The turning point occurs at $\upphi=0$ and satisfies
\begin{equation}\label{eq:cd-turning-condition}
 \tanh\uptau_\star
 =
 \frac{1-\alpha^2\delta^2}{1+\alpha^2\delta^2}
 \sec\frac{\alpha\phi_0}{2},
 \qquad
 \left.\frac{\diff\uptau}{\diff\upphi}\right|_{\uptau=\uptau_\star}=0\,.
\end{equation}
We choose $\uptau_\star$ with a positive real part, such that
\begin{align}\label{eq:cd-turning-points}
 \uptau_\star^{(n)}
 &=
 \operatorname{arctanh}\!\left[
 \frac{1+\alpha^2\delta^2}{1-\alpha^2\delta^2}
 \cos\frac{\alpha\phi_0}{2}
 \right]
 -\iu\pi\left(\frac12-n\right)
 \notag\\
 &=
 \log\left(\cot\frac{\alpha\phi_0}{4}\right)
 -\iu\pi\left(\frac12-n\right)
 +\mathcal O(\delta^2),
 \qquad n\in\mathbb Z.
\end{align}
The integer $n$ is the turning-point label introduced in
Eq.~\eqref{eq:turnP}. For every fixed $n$, the vacuum turning
point is recovered as $\alpha\to1$, and its asymptotic form agrees with
the vacuum expression when the cutoff is removed.

Substituting the embedding into
the global metric gives
\begin{equation}\label{eq:cd-induced}
 \diff\mathsf s_\gamma^2
 =
 -\Ls^2
 \frac{\cosh^2\uptau}
 {\sinh^2\uptau-\sinh^2\uptau_\star^{(n)}}
 \,\diff\uptau^2
 =
 -\frac{\Ls^2\,\diff u^2}{1+u^2},
 \qquad
 u=-\iu(-1)^n
 \frac{\sinh\uptau}{\sinh\uptau_\star^{(n)}}\,.
\end{equation}
We fix the length differential to be
$-\iu\Ls\,\diff u/\sqrt{1+u^2}$, choosing the square root positive at
the real UV endpoint and continuing it analytically into the bulk.
The imaginary UV divergence then has the sign of the $d=2$ term in
Eq.~\eqref{eq:Sexp}.

For this continuation, we use the contour defined in Eq.~\eqref{eq:contour-family} with $\tauUV=-\log(\alpha\delta)$ and the turning
point as given in Eq.~\eqref{eq:cd-turning-points},
using straight segments and fixing the sheet by analytic continuation
from the UV endpoint.\footnote{
The local equivalence of contours approaching the same ramification
point on different sheets, discussed in Sec.~\ref{subsec:higherdim_sphere} and Appendix~\ref{app:branch_contour},
does not eliminate nonzero global monodromies.}

Including the two symmetric halves of the extremal embedding, the area
functional becomes
\begin{equation}\label{eq:cd-area-integral}
 \operatorname{Area}_{\Gamma^{(n)}}(\Sigma_A)
 =
 2\iu\Ls
 \int_{\mathcal C_\uptau^{(n)}}\diff\uptau\,
 \frac{\cosh\uptau}
 {\sqrt{\sinh^2\uptau-\sinh^2\uptau_\star^{(n)}}}\,.
\end{equation}
The contour runs inward from the UV endpoint to $\uptau_\star^{(n)}$, with the
denominator square root positive at the real UV endpoint and analytically
continued along the contour. The factor of two is
$\operatorname{Vol}(\mathbb S^0)=2$ and accounts for the two halves of
the geodesic. For $\alpha=1$ and $\delta\to0$, the embedding reduces to Eq.~\eqref{eq:solUndef}, and the area
integral becomes the $d=2$ specialization of Eq.~\eqref{eq:AreaMaster}
with $T_0=\phi_0/2$.

Evaluating the integral on the sheet specified by
Eq.~\eqref{eq:contour-family}, we find
\begin{align}\label{eq:cd-length}
 \operatorname{Area}_{\Gamma^{(n)}}(\Sigma_A)
 &=
 (1-2n)\pi\Ls
 -2\iu\Ls\operatorname{arcosh}\!\left[
 \frac{1+\alpha^2\delta^2}{2\alpha\delta}
 \sin\frac{\alpha\phi_0}{2}
 \right]
 \nonumber\\
 &=
 (1-2n)\pi\Ls
 -2\iu\Ls
 \log\frac{\sin(\alpha\phi_0/2)}{\alpha\delta}
 +\mathcal O(\Ls\delta^2).
\end{align}
Increasing $n$ by one therefore shifts the area by
$-2\pi\Ls$ without changing its UV divergence. This exhibits the
global-period dependence discussed in Appendix~\ref{subsec:nshift} and makes explicit why
the continued length depends on the contour class as well as the
turning-point position.

Using the holomorphic RT prescription~\eqref{eq:RTformula}, we obtain
\begin{align}\label{eq:cd-entropy}
 S^{(n)}(A)
 &=
 \frac{(1-2n)\pi\Ls}{4\GN}
 -\frac{\iu\Ls}{2\GN}
 \log\frac{\sin(\alpha\phi_0/2)}{\alpha\delta}
 \nonumber\\
 &=
 \frac{c^{\rm dS}}{3}
 \left[
 \log\frac{\sin(\alpha\phi_0/2)}{\alpha\delta}
 +\iu\pi\left(\frac12-n\right)
 \right],
 \qquad
 c^{\rm dS}
 =12a^\star\big|_{d=2}
 =-\frac{3\iu\Ls}{2\GN}.
\end{align}
The central charge follows from the definition of $a^\star$ below
Eq.~\eqref{eq:Sdef1}. In the vacuum limit $\alpha\to1$, the logarithmic
term and universal coefficient reduce to the $d=2$ expressions in
Eq.~\eqref{eq:Sexp}, with $H=\sin(\phi_0/2)$ and
$\mathsf S_{\mathrm u}(\mathbb B^1)=4a^\star=c^{\rm dS}/3$.
The branch $n=0$ gives the HH contribution which can be directly compared with
Refs.~\cite{Anastasiou:2025rvz,Anastasiou:2026bbf}.

We now subtract the vacuum contribution on the same contour branch,
keeping the round boundary metric, endpoints, and cutoff fixed. Taking
$\delta\to0$ at fixed $\alpha$ and $\phi_0$ gives
\begin{equation}\label{eq:cd-state-difference}
 \Delta S(A)
 = S^{(n)}(A)-\left.S^{(n)}(A)\right|_{\alpha=1}
 =
 \frac{c^{\rm dS}}{3}
 \log\left[
 \frac{\sin(\alpha\phi_0/2)}
 {\alpha\sin(\phi_0/2)}
 \right]\,.
\end{equation}
Since the additive $n$-dependent contribution is independent of $\alpha$, it cancels in this
subtraction, and $\Delta S$ is independent of $n$ within the matched
contour family. The result vanishes both in the vacuum limit
$\alpha\to1$ and as the interval shrinks, i.e., $\phi_0\to0$. The factor
$1/\alpha$ is fixed by the cutoff relation~\eqref{eq:cd-global-cutoff}.

Writing $\mu=1-\alpha^2$ and expanding around the vacuum in Eq.~\eqref{eq:cd-state-difference} yields
\begin{equation}\label{eq:cd-first-order}
 \delta S(A)
 =
 \frac{c^{\rm dS}\mu}{6}
 \left[
 1-\frac{\phi_0}{2}\cot\frac{\phi_0}{2}
 \right]
 +\mathcal O(\mu^2)\,.
\end{equation}
Since we have expanded around the vacuum, we denote the entropy difference by $\delta$ instead of $\Delta$. This expression allows us to test the first law of pseudoentropy in the
linearized limit of the finite-defect result. Its holographic realization
through complex geodesics in dS$_3$ was demonstrated in
Ref.~\cite{Fujiki:2025rtx}.

We now study the variation of the modular Hamiltonian with respect to the vacuum. Let $\left.\tau_A\right|_{\alpha=1}$ be the normalized reduced transition
matrix of the reference vacuum, with modular Hamiltonian
\begin{equation}\label{eq:cd-modular-definition}
 K_A
 =-\log\left(\left.\tau_A\right|_{\alpha=1}\right)\,.
\end{equation}
For a normalized perturbation
$\tau_A=\left.\tau_A\right|_{\alpha=1}
+\delta\tau_A+\mathcal O(\mu^2)$ satisfying
$\operatorname{tr}_A\delta\tau_A=0$, the variation of
$S(A)=-\operatorname{tr}_A(\tau_A\log\tau_A)$ on a fixed analytic
branch of the logarithm gives
\begin{equation}\label{eq:cd-first-law}
 \delta S(A)
 =\operatorname{tr}_A(\delta\tau_A K_A)=
 \delta\langle K_A\rangle\,.
\end{equation}
Here $K_A$ remains fixed at its vacuum value, and the variation acts on
the transition matrix in its expectation value.

For a reference transition matrix supplied by the analytically continued CFT vacuum, the interval modular Hamiltonian retains
the standard local stress-tensor form continued to the nonunitary theory.
On the equator of the round boundary of radius $\Ls$, it reads
\begin{equation}\label{eq:cd-modular-hamiltonian}
 K_A
 =
 2\pi\Ls^2
 \int_{-\phi_0/2}^{\phi_0/2}\diff\phi\,
 \frac{\cos\phi-\cos(\phi_0/2)}{\sin(\phi_0/2)}
 T_{00}(\phi)
 +C_A\mathbf 1\,.
\end{equation}
The boundary energy density $T_{00}$ is defined in an orthonormal frame
obtained by analytically continuing the direction normal to the equator;
it is therefore distinct from an energy density associated with bulk
global time. The constant $C_A$ normalizes
$\left.\tau_A\right|_{\alpha=1}$ and drops out of
Eq.~\eqref{eq:cd-first-law}.

To compare the result of $\Delta S(A)$ with a semiclassical CFT state, we adopt the
proposed identification of a spinless defect with weights
\begin{equation}\label{eq:cd-weight}
 h=\bar h=\frac{c^{\rm dS}}{24}(1-\alpha^2)\,,
\end{equation}
with the corresponding analytically continued defect and uniformization
descriptions discussed in Refs.~\cite{Hikida:2022ltr,Asplund:2014coa}.
Under this identification, the semiclassical vacuum Virasoro block in the
heavy state reproduces Eq.~\eqref{eq:cd-state-difference}. The agreement
is at the level of this block and does not require it to dominate the complete nonunitary correlator.

Under the spinless-primary identification~\eqref{eq:cd-weight}, the
state--operator correspondence assigns an energy above the vacuum of
$(h+\bar h)/\Ls$ on a circle of circumference $2\pi\Ls$. The conformal
map from the round sphere to the radial-quantization cylinder has unit
Weyl factor at the equator, while its state-independent anomalous
contribution cancels in the vacuum subtraction. The resulting uniform
energy-density variation is thus
\begin{equation}\label{eq:cd-energy}
 \delta\langle T_{00}\rangle
 =
 \frac{h+\bar h}{2\pi\Ls^2}
 =
 \frac{c^{\rm dS}\mu}{24\pi\Ls^2}\,.
\end{equation}
Substituting it into the vacuum modular Hamiltonian gives
\begin{align}\label{eq:cd-modular-check}
 \delta\langle K_A\rangle
 &=
 2\pi\Ls^2
 \int_{-\phi_0/2}^{\phi_0/2}\diff\phi\,
 \frac{\cos\phi-\cos(\phi_0/2)}{\sin(\phi_0/2)}
 \,\delta\langle T_{00}\rangle
 \nonumber\\
 &=
 \frac{c^{\rm dS}\mu}{6}
 \left(
 1-\frac{\phi_0}{2}\cot\frac{\phi_0}{2}
 \right)
 =\delta S(A),
\end{align}
in agreement with Eq.~\eqref{eq:cd-first-order}. The first law is
therefore satisfied around the vacuum with the boundary metric, endpoints
and cutoff held fixed. Since the additive $n$-dependent contribution to the geodesic length is independent
of $\mu$, the check applies to every matched branch and does not select
$n$. In the short-interval limit, the state-dependent contribution becomes
\begin{equation}
 \delta S(A)
 =
 \frac{c^{\rm dS}\mu\phi_0^2}{72}
 +\mathcal O(\mu\phi_0^4,\mu^2)\,.
\end{equation}
This calculation provides a nonvacuum test of the holomorphic prescription, such that the first law of pseudoentropy is satisfied independently of the homology class (labeled by $n$), provided that the same class is chosen for both the vacuum and the excited state. Thus, the integration cycle degeneracy is not lifted for this class of excited states.
 
\section{Discussion and conclusions}\label{sec:conclusions}
 
In this work, we have developed a holomorphic extremization framework for computing holographic pseudoentropy from codimension-two extremal surfaces in complexified dS geometry, while imposing real boundary conditions at future infinity by specifying the entangling region of the dual nonunitary CFT. The extremization problem and its associated equation of motion ($K^I=0$) can be formulated directly in terms of holomorphic bulk coordinates, so that the embedding of the extremal surface is obtained without introducing separate Lorentzian and Euclidean pieces from the start. For the spherical regions and their harmonic deformations considered here, the resulting embedding and area functional are holomorphic within their relevant domains. Consequently, whenever two integration contours have the same endpoints and can be continuously deformed into one another without crossing singularities or branch cuts, Cauchy's theorem ensures that they give the same extremal area. The holographic pseudoentropy is therefore independent of the particular representative within such a class of admissible contours. For $n=0$, this reproduces the result of the piecewise HH construction for unit-size spheres and their deformations, and extends it to arbitrary size, where no real piecewise surface exists, and to arbitrary dimensions. Different turning-point labels $n$ correspond to inequivalent relative cycles on the same complex surface. Their areas differ by an overall sign of the finite part in odd $d$ and by real periods in even $d$, a structure that follows from the $\uptau\to\uptau+\iu\pi$ symmetry.
 
It is important, however, to distinguish this contour equivalence at the level of the area from an equivalence of the corresponding geometric bulk realizations. All the contours considered here are integration cycles within the same holomorphically complexified dS manifold, but they need not select the same section of this manifold. This distinction is particularly transparent for contours composed of purely real and purely imaginary segments. A real segment of the global time coordinate selects a Lorentzian dS section, whereas a purely imaginary segment selects a Euclidean dS section. A piecewise contour formed from successive segments of this type therefore admits a direct interpretation as a piecewise bulk geometry obtained by gluing Lorentzian and Euclidean dS regions, as in the usual HH prescription~\cite{Skenderis:2008dh}. Changing the contour can consequently change this piecewise geometric realization even though the corresponding contours remain continuously deformable in the full complexified geometry. For a generic contour with simultaneously varying real and imaginary parts, the induced geometry is instead genuinely complex and does not admit an interpretation as an ordinary Lorentzian or Euclidean spacetime. Therefore, for the geometric vacuum, even though the pseudoentropy does not change within a class of different geometric realizations embedded in the same complexified bulk, not all admissible contours describe the same bulk geometry.
 
We emphasize that this contour independence is a property of the geometric vacuum considered here and should not be expected to hold for arbitrary states or bulk backgrounds. Already for perturbations around pure dS, complexified extremal surfaces play a non-trivial role in recovering the first law of pseudoentropy and, at linear order, the bulk gravitational equations of motion~\cite{Fujiki:2025rtx}. In Sec.~\ref{sec:ds3-conical-defect}, we explicitly consider a restricted type of metric deformation around the dS vacuum consisting of introducing a conical defect by modifying the angular periodicity of the transverse section in global coordinates. In this case, we see that the complex analysis for computing the geodesic length (codimension-2 area in the corresponding dS$_3$/CFT$_2$ case) already gives compatibility with the first law of pseudoentropy while still allowing for the contour degeneracy among the representative sections of the complex RT geodesic. Therefore, the degeneracy may persist even for perturbed dS geometry, although the deformation is still around the geometric vacuum.
An even clearer distinction arises in black-hole backgrounds. Complex codimension-two extremal surfaces have been explicitly identified in BTZ, Schwarzschild-AdS and Schwarzschild-Lifshitz geometries~\cite{Fischetti:2014zja}, and studies of timelike entanglement in black-hole spacetimes show that several complex saddles may contribute and that their behavior can be sensitive to the presence of the black-hole singularity~\cite{Anegawa:2024kdj,Heller:2024whi,Heller:2025kvp}. In such backgrounds, the analytic continuation of the extremization problem develops a considerably richer structure: the black-hole metric functions introduce additional zeros and poles, while the extremal embedding and its area can develop turning points, branch points and associated branch cuts. Contours that lie on different sheets or cannot be deformed into one another without crossing these obstructions need not yield the same area. In this case, the distinction between complex saddles is therefore physical at the semiclassical level rather than merely a freedom in choosing a contour representative.
 
A related situation is expected for black holes in dS. Codimension-two extremal surfaces in Schwarzschild-dS already display a non-trivial sensitivity to the cosmological and black-hole horizons and to the extended causal structure of the spacetime~\cite{Fernandes:2019sds}, while complex extremal surfaces in deformations of pure dS exhibit the richer behavior expected once the analytic structure of the geometric vacuum is modified~\cite{Narayan:2015vda}. It would therefore be particularly interesting to apply the holomorphic formulation developed here directly to Schwarzschild-dS and related backgrounds. In contrast with pure dS, inequivalent integration cycles may then explore different analytic branches of the codimension-two extremization problem and yield distinct values of the area. In this sense, the breakdown of the contour equivalence found here could itself become a useful probe of the analytic structure associated with horizons and the black-hole interior. A systematic investigation of this possibility lies beyond the scope of the present work.
 
The contour independence of the geometric vacuum also has a broader implication for the interpretation of the holographic bulk. From the point of view of the pseudo-entanglement structure of the boundary state, the different admissible geometric realizations described above cannot be distinguished by their classical extremal areas. In particular, the usual HH geometry---with its specific gluing of Euclidean and Lorentzian dS sections---is only one representative within a larger family of complex contour realizations that reproduce the same pseudoentropy. Since the boundary conditions are fixed on the real conformal boundary while the continuation into the complexified bulk is not uniquely selected by this observable, the present result supports the interpretation of the bulk geometry as an emergent or auxiliary representation of the underlying boundary data. More precisely, the pseudoentropy determines, at this level, an equivalence class of bulk geometric realizations rather than selecting a unique member of that class. This does not imply that the different contours define distinct complexified manifolds---they are sections or cycles of the same complexified dS geometry---but it does mean that geometrically different realizations of the bulk can be indistinguishable from the perspective of the boundary quantity computed here.
 
There are several ways in which this degeneracy may cease to hold. Perturbations away from the geometric vacuum can introduce state-dependent contributions that distinguish complex extremal configurations~\cite{Fujiki:2025rtx}. Black-hole backgrounds provide a classical mechanism by modifying the analytic structure of the extremization problem and allowing inequivalent branches of the complexified surface~\cite{Fischetti:2014zja,Anegawa:2024kdj}. Quantum corrections provide a further possibility~\cite{Faulkner:2013ana,Barrella:2013wja,Engelhardt:2014gca,Jafferis:2015del,Agon:2015ftl,Dong:2017xht,Agon:2026irg,Belin:2018juv}. At the semiclassical level considered here, different contour realizations are compared through their classical complex areas. Once the RT prescription is replaced by an appropriate complexified quantum extremal surface prescription, the generalized entropy would also contain the entanglement of bulk quantum fields across the surface. Two configurations with the same classical area need not have the same bulk entanglement contribution and could therefore become distinguishable beyond the classical approximation. Understanding whether such corrections select a preferred representative within the class of complex bulk geometries would provide an important extension of the present analysis.
 
Overall, our results show that for the geometric vacuum of dS/CFT the usual piecewise HH construction can be embedded into a single holomorphic codimension-two extremization problem. Different integration contours may correspond to different geometric realizations of the bulk---with the distinction being particularly concrete when the contour joins Euclidean and Lorentzian dS sections---while nevertheless yielding the same holographic pseudoentropy whenever they are continuously deformable within a common analytic domain. The absence of a preferred realization at this level is consistent with regarding the bulk geometry as an emergent description determined only up to such an equivalence. At the same time, the richer analytic structure of excited states, black-hole backgrounds and quantum-corrected extremal surfaces provides natural mechanisms by which this equivalence may be broken, allowing genuinely distinct complex saddles to have different complex areas for their extremal surfaces and therefore different values of pseudoentropy.

\section*{Acknowledgements}
We thank Wu-zhong Guo and Tadashi Takayanagi for interesting discussions. This work is partially funded by ANID FONDECYT grants 11240059, 1240043, 1261016 and 1231133. I.J.A. and A.D. gratefully acknowledge support from the Yukawa Institute for Theoretical Physics, Kyoto University, where some of the research for this work was performed. The work of A.D. is supported by Becas de postgrado UNAP. J.M. is supported by the European Union's Horizon Europe research and innovation programme under the Marie Sk\l odowska-Curie grant agreement
No.~101202710 (QUBITTO).
 
\appendix
 
\section{Geometric interpretation of extremality condition}
\label{app:TrK}
 
We briefly derive the equivalence between the extremality of the area functional and the vanishing of the trace of the extrinsic curvature appearing in \eqref{eq:ExtCon}. Let the codimension-two surface $\Sigma_A$ be parametrized by an embedding
\begin{equation}
X^\mu = X^\mu(\sigma^a)\,,
\qquad
a=1,\ldots,d-1\,,
\end{equation}
with tangent vectors
\begin{equation}
e_a^\mu=\partial_a X^\mu\,.
\end{equation}
The induced metric on $\Sigma_A$ is
\begin{equation}
\gamma_{ab}
=
g_{\mu\nu}e_a^\mu e_b^\nu\,,
\label{eq:app_induced_metric}
\end{equation}
and the area functional evaluated on the integration cycle $\Gamma$ is given by
\begin{equation}
\Area=\int_{\Gamma}\diff^{d-1}\sigma\,\sqrt{\gamma}\,.
\end{equation} 
All quantities in this appendix are understood as complex-valued in the complexified bulk.
 
An infinitesimal deformation of the embedding can be decomposed into tangential and normal components,
\begin{equation}
\delta X^\mu
=
\xi^a e_a^\mu
+
\Phi^I n_I^\mu,
\label{eq:app_deformation}
\end{equation}
where $n_I^\mu$, with $I=1,2$, are the two independent normal vectors satisfying
\begin{equation}
g_{\mu\nu}n_I^\mu e_a^\nu = 0\,.
\end{equation}
The tangential deformation $\xi^a e_a^\mu$ corresponds to a reparametrization of the surface and does not change its geometric area. It is therefore sufficient to consider the normal deformation
\begin{equation}
\delta X^\mu = \Phi^I n_I^\mu\,.
\end{equation}
 
The covariant derivative of the tangent vectors admits the Gauss decomposition
\begin{equation}
\nabla_a e_b^\mu
=
\Gamma^c{}_{ab} e_c^\mu
+
K^I{}_{ab} n_I^\mu\,,
\label{eq:app_Gauss}
\end{equation}
where
\begin{equation}
K^I{}_{ab}
=
n^I{}_\mu \nabla_a e_b^\mu
\end{equation}
is the extrinsic curvature associated with the normal direction $n_I^\mu$. Under a normal deformation, the induced metric varies as
\begin{equation}
\delta\gamma_{ab}
=
2\Phi^I K^I{}_{ab}\,,
\label{eq:app_delta_gamma}
\end{equation}
where the overall sign depends on the convention chosen for the extrinsic curvature. Consequently,
\begin{equation}
\delta\Area=
\frac{1}{2}
\int_{\Gamma} \diff^{d-1}\sigma\,
\sqrt{\gamma}\,
\gamma^{ab}\delta\gamma_{ab}=
\int_{\Gamma} \diff^{d-1}\sigma\,
\sqrt{\gamma}\,
\Phi^I \gamma^{ab}K^I{}_{ab}\,.
\label{eq:app_delta_area}
\end{equation}
Since the normal deformations $\Phi^I$ are arbitrary, stationarity of the area functional
\begin{equation}
\delta \Area=0
\end{equation}
requires
\begin{equation}
\gamma^{ab}K^I{}_{ab}=0\,,
\qquad
I=1,2\,.
\label{eq:app_TrK}
\end{equation}
Thus, the extremality condition is equivalent to the vanishing of the mean curvature vector of $\Sigma_A$. In the notation of \eqref{eq:ExtCon}, this is simply
\begin{equation}
K^I=0\,.
\end{equation}
 
In the complexified geometry considered here, the same argument applies with the induced metric and extrinsic curvature understood as complex quantities. The extremal surface is therefore characterized by the vanishing of its complex mean curvature vector, i.e.
\begin{equation}
K^{I}_{\mathbb{C}}=0\,.
\label{eq:app_complex_TrK}
\end{equation}
This establishes the equivalence between the variational condition used to define the holomorphic extremal surface and the geometric condition stated in \eqref{eq:ExtCon}.
 
\section{Origin of the $\sin T_0$ factor from stereographic projection}
\label{app:stereographic}
 
In this appendix we derive the normalization of the deformation appearing in
Eq.~\eqref{eq:phiF}. The factor $\sin T_0$ is not introduced by hand, but
follows directly from the stereographic projection relating the Euclidean plane
to the boundary sphere.
 
Consider the stereographic map
\begin{equation}
r=\tan\frac{\tE}{2}\,,
\label{eq:stereo}
\end{equation}
where $r$ is the radial coordinate on the plane and $\tE$ is the polar angle on
the boundary sphere.
 
Suppose the undeformed entangling surface is a circle of radius $r_0$,
corresponding to
\begin{equation}
r_0=\tan\frac{T_0}{2}\,,
\qquad
T_0=2\arctan r_0\,.
\end{equation}
We now introduce a small harmonic deformation in the plane,
\begin{equation}
r(\phi)=r_0\left(1+\epsilon\,f(\phi)\right),
\qquad
\epsilon\ll1\,,
\end{equation}
where
\begin{equation}
f(\phi)
=
\sum_{\ell}
\left[
a_\ell\cos(\ell\phi)
+
b_\ell\sin(\ell\phi)
\right].
\end{equation}
 
Using the inverse stereographic map,
\begin{equation}
\tE(\phi)
=
2\arctan\!\left[r_0\left(1+\epsilon f(\phi)\right)\right]\,,
\end{equation}
We expand to first order in $\epsilon$. Since
\begin{equation}
\frac{\diff}{\diff x}\arctan x=\frac{1}{1+x^2}\,,
\end{equation}
the Taylor expansion gives
\begin{equation}
\arctan(x_0+\delta x)
=
\arctan x_0
+
\frac{\delta x}{1+x_0^2}
+
\mathcal O(\delta x^2)\,,
\end{equation}
with
\begin{equation}
x_0=r_0,
\qquad
\delta x=\epsilon r_0 f(\phi)\,.
\end{equation}
Hence,
\begin{equation}
\tE(\phi)
=
2\arctan r_0
+
\epsilon
\frac{2r_0}{1+r_0^2}
f(\phi)
+
\mathcal O(\epsilon^2)\,.
\end{equation}
 
Finally, using
\begin{equation}
r_0=\tan\frac{T_0}{2}\,,
\end{equation}
we obtain
\begin{equation}
\frac{2r_0}{1+r_0^2}
=
\frac{2\tan(T_0/2)}
{1+\tan^2(T_0/2)}
=
2\sin\frac{T_0}{2}\cos\frac{T_0}{2}
=
\sin T_0.
\end{equation}
Therefore,
\begin{equation}
\tE(\phi)
=
T_0
+
\epsilon\,\sin T_0\,f(\phi)
+
\mathcal O(\epsilon^2)\,.
\end{equation}
 
This is precisely the normalization adopted in Eq.~\eqref{eq:phiF}. Thus, the
factor $\sin T_0$ is completely determined by the stereographic projection from
the plane to the boundary sphere and is therefore geometrically fixed rather
than a matter of convention.
 
\section{Contour continuation around the turning point}
\label{app:branch_contour}
 
In this appendix we make explicit the relation between contours that differ
by a winding around a turning point in even dimensions. We focus on
$d=4$, for which the area functional following from
Eq.~\eqref{eq:AreaMaster} takes the form
\begin{equation}
   \Area=
    4\pi\iu\,\Ls^3\sin T_0
    \int_{\mathcal C_\uptau}
    \diff\uptau\,
    \cosh\uptau
    \sqrt{1+\sin^2T_0\sinh^2\uptau}\,.
\label{eq:AppAreaD4}
\end{equation}
The square root gives rise to branch points at the turning points of the
extremal surface. We show that, when a contour winds once around a single turning
point and subsequently terminates at that same turning point, the resulting
area agrees exactly with that of the direct contour.
 
For this, it is useful first to introduce
\begin{equation}
    s=\sin T_0,
    \qquad
    u=\sinh\uptau,
    \qquad
    \diff u=\cosh\uptau\,\diff\uptau.
\end{equation}
The area integral then becomes
\begin{equation}
    \mathcal I[\mathcal C]
    =
    \int_{\mathcal C_u}\diff u\,
    \sqrt{1+s^2u^2}\,,
\label{eq:AppIu}
\end{equation}
up to the overall prefactor in Eq.~\eqref{eq:AppAreaD4}. The two branch
points in the $u$-plane are
\begin{equation}
    u_\pm=\pm\frac{\iu}{s}\,.
\label{eq:AppBranchPointsU}
\end{equation}
 
Consider the turning point
\begin{equation}
    \uptau_\star
    =
    \log\!\left(\cot\frac{T_0}{2}\right)
    -\frac{\iu\pi}{2}\,.
\label{eq:AppTauStar}
\end{equation}
Using
\begin{equation}
    \sinh(a-\iu\pi/2)
    =
    -\iu\cosh a,
    \qquad
    a=\log\!\left(\cot\frac{T_0}{2}\right)\,,
\end{equation}
together with
\begin{equation}
    \cosh\!\left[
    \log\left(\cot\frac{T_0}{2}\right)
    \right]
    =
    \frac{1}{\sin T_0}\,,
\end{equation}
we obtain
\begin{equation}
    u_\star=-\frac{\iu}{s}\,.
\label{eq:AppUStar}
\end{equation}
Thus the turning point is precisely one of the branch points of
Eq.~\eqref{eq:AppIu}.
 
Near this point, introduce the local uniformizing coordinate $w$ by
\begin{equation}
    u-u_\star=w^2\,.
\label{eq:AppUniformizing}
\end{equation}
The two sheets of the square root are then represented by the two signs
of $w$, while the branch point itself corresponds to the single point
$w=0$.
 
Using $u_\star=-\iu/s$, we have
\begin{align}
    1+s^2u^2
    &=
    1+s^2
    \left(
        -\frac{\iu}{s}+w^2
    \right)^2
    \nonumber\\
    &=
    1-1-2\iu s\,w^2+s^2w^4
    \nonumber\\
    &=
    w^2\left(-2\iu s+s^2w^2\right)\,.
\label{eq:AppSquareRootExpansion}
\end{align}
Hence, on a chosen local branch,
\begin{equation}
    \sqrt{1+s^2u^2}
    =
    w\sqrt{-2\iu s+s^2w^2}\,,
\label{eq:AppSqrtW}
\end{equation}
where the sign of $w$ keeps track of the sheet. Since
\begin{equation}
    \diff u=2w\,\diff w\,,
\end{equation}
the area differential becomes
\begin{equation}
    \Omega
    =
    \sqrt{1+s^2u^2}\,\diff u
    =
    2w^2
    \sqrt{-2\iu s+s^2w^2}\,\diff w\,.
\label{eq:AppOmegaW}
\end{equation}
The important point is that $\Omega$ is holomorphic at $w=0$. The apparent
square-root branch singularity in the $u$-plane has therefore disappeared
in the uniformizing coordinate.
 
Now, let $\uptau_r$ be a point close to $\uptau_\star$, and define
\begin{equation}
    u_r-u_\star=w_r^2\,,
    \qquad
    w_r=r\,,
    \qquad
    r>0\,.
\label{eq:AppRegulator}
\end{equation}
The direct contour from $\uptau_r$ to $\uptau_\star$ corresponds
locally to $w:r\longrightarrow0.$
Its contribution is therefore
\begin{equation}
    \mathcal I_{\rm direct}
    =
    \int_r^0 \Omega\,.
\label{eq:AppDirect}
\end{equation}
 
Now consider instead a contour which reaches $\uptau_r$ on the
first sheet, winds once around the branch point $\uptau_\star$, and then approaches the
turning point on the second sheet. A single winding in the $u$-plane
corresponds to
\begin{equation}
    w:r\longrightarrow-r\,.
\label{eq:AppWinding}
\end{equation}
After one winding, the contour
has moved from one sheet to the other.
 
The subsequent segment from $\uptau_r$ to $\uptau_\star$ on the
second sheet corresponds to
\begin{equation}
    w:-r\longrightarrow0\,.
\end{equation}
Thus the combined contribution of the winding and second-sheet segments
is
\begin{equation}
    \mathcal I_{\rm wind+II}
    =
    \int_r^{-r}\Omega
    +
    \int_{-r}^{0}\Omega\,.
\label{eq:AppWindSecond}
\end{equation}
Since $\Omega$ is a single-valued holomorphic differential in the
uniformizing coordinate $w$, the two consecutive integrals combine
exactly:
\begin{equation}
    \mathcal I_{\rm wind+II}
    =
    \int_r^0\Omega
    =
    \mathcal I_{\rm direct}\,.
\label{eq:AppExactEquality}
\end{equation}
Therefore,
\begin{equation}
    \mathcal I_{\rm wind+II}
    -
    \mathcal I_{\rm direct}
    =0
\label{eq:AppZeroDifference}
\end{equation}
for any sufficiently small but finite $r$.
 
The common ultraviolet portion of the two contours is identical. Hence
the full area functionals obey
\begin{equation}
    \mathcal A_{\rm wind+II}
    =
    \mathcal A_{\rm direct}\,.
\label{eq:AppAreaEquality}
\end{equation}
 
\subsection*{Explicit small-$r$ behavior}
 
It is also useful to examine the behavior of the individual local pieces.
From Eq.~\eqref{eq:AppOmegaW},
\begin{equation}
    \Omega
    =
    2\sqrt{-2\iu s}\,w^2\,\diff w
    +O(w^4)\,\diff w\,.
\label{eq:AppOmegaExpansion}
\end{equation}
Consequently, the direct segment behaves as
\begin{align}
    \mathcal I_{\rm direct}
    &=
    2\sqrt{-2\iu s}
    \int_r^0w^2\,\diff w
    +O(r^5)
    \nonumber\\
    &=
    -\frac{2}{3}\sqrt{-2\iu s}\,r^3
    +O(r^5)\,.
\label{eq:AppDirectScaling}
\end{align}

Similarly, the winding segment satisfies
\begin{align}
    \mathcal I_{\rm winding}
    &=
    2\sqrt{-2\iu s}
    \int_r^{-r}w^2\,\diff w
    +O(r^5)
    \nonumber\\
    &=
    -\frac{4}{3}\sqrt{-2\iu s}\,r^3+O(r^5),
\label{eq:AppWindingLeading}
\end{align}
where the integral may be evaluated along any path from $w=r$ to $w=-r$, e.g.\ the semicircle traced by the winding. The winding piece is thus nonzero at finite $r$,
and is suppressed by positive powers of $r$, so that it vanishes as
$r\to0$.
 
The second-sheet segment similarly gives
\begin{align}
    \mathcal I_{\rm II}
    &=
    2\sqrt{-2\iu s}
    \int_{-r}^{0}w^2\,\diff w
    +O(r^5)
    \nonumber\\
    &=
    +\frac{2}{3}\sqrt{-2\iu s}\,r^3
    +O(r^5).
\label{eq:AppSecondScaling}
\end{align}
 
The individual local pieces therefore vanish in the limit
$r\to0$. More importantly, however, their combination with the
appropriate orientations reproduces the direct contour exactly at finite
$r$, as shown in Eq.~\eqref{eq:AppExactEquality}.
 
The result can be summarized geometrically. In the complex $\uptau$-plane,
the turning point appears as a square-root branch point, and a loop around
it appears to return to the same value of $\uptau$. On the Riemann surface,
however, the lifted path is an open path connecting the two sheets. The
local coordinate $w$ makes this structure manifest:
\begin{equation}
    \text{sheet I}:w>0,
    \qquad
    \text{sheet II}:w<0,
    \qquad
    \uptau=\uptau_\star\Longleftrightarrow w=0\,.
\end{equation}
The winding segment is therefore simply the path $r\to-r$, followed by
the second-sheet segment $-r\to0$. Their concatenation is the direct path
$r\to0$.
 
Thus, changing sheets by winding around a turning point does not by itself
produce a new finite contribution to the area when the contour ultimately
terminates at the same ramification point. Genuine contour dependence can
arise only when the analytic continuation changes the global endpoint
structure or connects to a different part of the Riemann surface, for instance by winding around the $\uptau$-cylinder, which shifts the area by an n-dependent monodromy (see Appendix~\ref{subsec:nshift}). The
absence of an additional contribution in the present case is therefore a
local statement about the regularity of the area differential in the
uniformizing coordinate, rather than a consequence of the arbitrary choice
of branch cut in the complex $\uptau$-plane.

\section{Turning-point regularity of the quadratic area integral}
\label{app:turningpoint}

We briefly analyze the behavior of the shape-deformation modes at the
turning point of the spherical extremal surface. The purpose is to verify
that the regular branch selected in the main text does not introduce a
singularity in the quadratic area integral. This justifies the contour
deformations used within a fixed relative homology class.

For convenience, define
\begin{equation}
\chi(\uptau)
=
1+\sin^2 T_0\,\sinh^2\uptau \,,
\qquad
\chi(\uptau_\star)=0 \,.
\label{eq:app_chi}
\end{equation}
The turning points are therefore precisely the zeros of $\chi$. We use
\begin{equation}
q=\sqrt{\chi}
\label{eq:app_q}
\end{equation}
as a local coordinate on the two-sheeted surface obtained by resolving
the square-root behavior at the turning point.
For simplicity, in the following local analysis we set
$T_0=\frac{\pi}{2}.$ This choice does not affect the conclusion of the analysis as for a general $T_0$, the turning point is still a regular singular point with the same local structure and the corresponding expressions are related by a nonsingular rescaling of the local coordinate.

Therefore, the radial equation for a harmonic mode can be written locally near
$\uptau_\star$ as
\begin{equation}
q^2\frac{\diff^2\mathcal T_\ell}{\diff q^2}
+
d\,q\frac{\diff\mathcal T_\ell}{\diff q}
-
c_\ell\,\mathcal T_\ell
=
\mathcal O(q)\,,
\label{eq:app_local_mode}
\end{equation}
where
\begin{equation}
c_\ell
=
\ell^2+(d-3)\ell-(d-2)
=
(\ell-1)(\ell+d-2).
\label{eq:app_cell}
\end{equation}
Assuming a behavior given by $\mathcal T_\ell
\sim
q^{p}$ near the $q=0$ turning point, one obtains the algebraic equation 
\begin{equation}
p(p-1)+dp-c_\ell=0\,,
\label{eq:app_indicial}
\end{equation}
with roots
\begin{equation}
p_+=\ell-1,
\qquad
p_-=-(\ell+d-2)\,.
\label{eq:app_roots}
\end{equation}
Equivalently, in terms of $\chi$ the two local behaviors are
\begin{equation}
\mathcal T_\ell
\sim
\chi^{(\ell-1)/2}\,,
\qquad
\mathcal T_\ell
\sim
\chi^{-(\ell+d-2)/2}\,.
\label{eq:app_chi_branches}
\end{equation}
The boundary-value problem considered in the main text selects the branch that is regular at $\uptau_\star$. 
We therefore restrict the following local analysis to this regular branch. The other Frobenius branch is singular at $\uptau_\star$ and is excluded by the regularity condition imposed on the extremal surface. Our purpose here is only to verify that the regular branch selected in the main text is harmless under local contour deformations around the turning point. Thus, for $\ell\geq2$,
\begin{equation}
\mathcal T_\ell(q)
=
b_0\,q^{\ell-1}
\left(1+\mathcal O(q)\right)\,,
\label{eq:app_regular_branch}
\end{equation}
and consequently
\begin{equation}
\mathcal T_\ell(\uptau_\star)=0,
\qquad
\ell\geq2\,.
\label{eq:app_Tzero}
\end{equation}

We now examine the quadratic area contribution. After the angular
integration, orthogonality of the spherical harmonics reduces the
shape-dependent term to a sum of independent radial integrals,
\begin{equation}
\mathcal A^{(2)}
=
\frac{\sigma L^{d-1}\epsilon^2}{2}
\sum_{\ell,\mathbf m}
a_{\ell,\mathbf m}^2
\int_{\mathcal C_\uptau}
\diff\uptau\,\mathcal I_\ell(\uptau)\,,
\label{eq:app_A2}
\end{equation}
where $\mathcal I_\ell$ denotes the quadratic radial integrand for a given
mode. Near the turning point, its leading dependence on the local
coordinate $q$ takes the form
\begin{equation}
\mathcal I_\ell
=
q^{d-2}c_\ell\mathcal T_\ell^2
+
q^d
\left(
\frac{\diff\mathcal T_\ell}{\diff q}
\right)^2
+
\mathcal O(q^{2\ell+d-3})\,.
\label{eq:app_Ilocal}
\end{equation}
Using the regular solution \eqref{eq:app_regular_branch}, we obtain
\begin{align}
q^{d-2}c_\ell\mathcal T_\ell^2
&=
c_\ell b_0^2\,
q^{2\ell+d-4}
\left(1+\mathcal O(q)\right)\,,
\label{eq:app_I1}
\\
q^d
\left(
\frac{\diff\mathcal T_\ell}{\diff q}
\right)^2
&=
(\ell-1)^2 b_0^2\,
q^{2\ell+d-4}
\left(1+\mathcal O(q)\right)\,.
\label{eq:app_I2}
\end{align}
Hence
\begin{equation}
\mathcal I_\ell
=
(2\ell+d-3)(\ell-1)b_0^2\,
q^{2\ell+d-4}
\left(1+\mathcal O(q)\right)\,.
\label{eq:app_Iregular}
\end{equation}

For $T_0\neq\pi/2$, $\chi$ has a simple zero at $\uptau_\star$, so that
$\diff\uptau=\mathcal O(q)\diff q$. Therefore, the quadratic area
one-form behaves as
\begin{equation}
\mathcal I_\ell(\uptau)\diff\uptau
=
\mathcal O\!\left(q^{2\ell+d-3}\right)\diff q\,,
\qquad
q\rightarrow0\,.
\label{eq:app_area_oneform}
\end{equation}
It is consequently regular at $q=0$ and contains no $q^{-1}\diff q$
term. In particular,
\begin{equation}
\operatorname*{Res}_{q=0}
\left[
\mathcal I_\ell(\uptau)\diff\uptau
\right]
=0\,.
\label{eq:app_residue}
\end{equation}

At the special point $T_0=\pi/2$, one has
$\chi=\cosh^2\uptau$ and
$q\sim\uptau-\uptau_\star$, so that the same conclusion follows
directly from
\begin{equation}
\mathcal I_\ell(\uptau)
=
\mathcal O\!\left(
(\uptau-\uptau_\star)^{2\ell+d-4}
\right)\,.
\label{eq:app_unit_radius}
\end{equation}

The same result can be seen directly from the on-shell form of the
quadratic functional. Using the leading radial equation for the harmonic mode
\eqref{eq:app_local_mode}, the local integrand becomes a total
derivative,
\begin{equation}
\mathcal I_\ell
=
\frac{\diff}{\diff q}
\left[
q^d\mathcal T_\ell
\frac{\diff\mathcal T_\ell}{\diff q}
\right]
+
\mathcal O(q^{2\ell+d-3})\,.
\label{eq:app_total_derivative}
\end{equation}
For the regular branch,
\begin{equation}
q^d\mathcal T_\ell
\frac{\diff\mathcal T_\ell}{\diff q}
=
(\ell-1)b_0^2\,
q^{2\ell+d-3}
\left(1+\mathcal O(q)\right)\,,
\label{eq:app_boundary_term}
\end{equation}
which vanishes at $q=0$ for every $\ell\geq2$. Thus the turning point
does not contribute a boundary term to the on-shell quadratic area.

We therefore conclude that the turning point is locally harmless for
the regular deformation saddle: although it can appear as a branch
point in the $\uptau$-plane when the area integrand itself is
multivalued, the quadratic area differential is regular on the
corresponding uniformized surface. Consequently, two contours ending
at the same turning point and related by a local deformation around it
give the same quadratic area, provided no other singularity or
nontrivial global cycle---labeled by n in the main text---is crossed. This is precisely the contour
statement needed in the main text. Note that contours within a fixed homology
class are equivalent, whereas contours belonging to different classes
can differ by the global monodromies associated with the
$\uptau\rightarrow\uptau+\iu\pi$ shift. Those global effects are
responsible for the $n$-dependence of the universal terms discussed in
Appendix~\ref{subsec:nshift}.

\section{Contour classes organized by n-winding on the $\uptau$ cylinder}
\label{subsec:nshift}

The dependence on $n$ follows from the shift
$\uptau\to\uptau+\iu\pi$ underlying Eq.~\eqref{eq:Gamma-n}.
The spherical profile~\eqref{eq:solUndef} is invariant under
this shift, while the analytically continued area integrand $\mathcal{L}$ satisfies
\begin{equation}\label{eq:Lshift}
\mathcal L(\uptau+\iu\pi)=(-1)^d\,\mathcal L(\uptau)\,.
\end{equation}
The linearized extremality equation~\eqref{eq:diffeqHigherD}
is also invariant. Let $\mathcal T^{(n)}_\ell$ be its
solution approaching unity at the conformal boundary and regular
at $\uptau_\star^{(n)}$. Whenever these conditions select a unique
solution, they imply
\begin{equation}\label{eq:Tshift}
\mathcal T^{(n+1)}_\ell(\uptau)
=
\mathcal T^{(n)}_\ell(\uptau-\iu\pi)\,,
\qquad
\uptau_\star^{(n+1)}=\uptau_\star^{(n)}+\iu\pi\,.
\end{equation}
Thus, the shift also relates the perturbative surfaces used
in the deformation calculations.

As in Eq.~\eqref{eq:Gamma-n}, the shifted contour is supplemented
by a UV segment from $\tauUV$ to $\tauUV+\iu\pi$.
The argument leading to Eq.~\eqref{eq:Sexp} then applies:
the power-law divergences are independent of $n$, while the
finite term in odd $d$ acquires the factor $(-1)^n$.
In even $d$, each shift instead replaces
$\log(H/\delta)$ by $\log(H/\delta)-\iu\pi$.

\bibliographystyle{JHEP}
\bibliography{Biblio}
 
\end{document}